\documentclass[11pt,twoside,a4paper,cmspaper,final,collab]{cms-tdr}
\def\svnVersion{8209bcb-D}\def\svnDate{2026/06/08}\def\cmsCernNoTag{CERN-EP-2026-149}\def\cmsCernDate{\today}\def\cmsMessage{Submitted to the Journal of High Energy Physics}
\begin{document}\cmsNoteHeader{B2G-25-004}

\newcommand{\bbtt}{\ensuremath{\PQb\PQb\PGt\PGt}\xspace}
\newcommand{\tautau}{\ensuremath{\PGt\PGt}\xspace}
\newcommand{\bb}{\ensuremath{\PQb\PQb}\xspace}

\newcommand{\taue}{\HepParticle{\PGt}{\!\Pe}{}\xspace}
\newcommand{\taumu}{\HepParticle{\PGt}{\!\PGm}{}\xspace}
\providecommand{\PWst}{\HepParticle{\PW}{\ast}\xspace}
\providecommand{\PGgst}{{\HepParticle{\PGg}{}{\ast}}\xspace}
\newcommand{\muth}{\ensuremath{\taumu\tauh}\xspace}
\newcommand{\eleth}{\ensuremath{\taue\tauh}\xspace}
\newcommand{\thth}{\ensuremath{\tauh\tauh}\xspace} 

\newcommand{\ZHbb}{\ensuremath{\PZ/\PH\to \PQb\PQb}\xspace}
\newcommand{\ZHtt}{\texorpdfstring{\ensuremath{\PZ/\PH\to \PGt\PGt}\xspace}{Z/H to tautau}\xspace}

\newcommand{\ZZ}{\ensuremath{\PZ\PZ}\xspace}
\newcommand{\ZH}{\ensuremath{\PZ\PH}\xspace}
\newcommand{\zbbhtt}{\ensuremath{\HepParticle{\PZ}{\PQb\PQb}{}\HepParticle{\PH}{\PGt\PGt}{}}\xspace}
\newcommand{\ztthbb}{\ensuremath{\HepParticle{\PZ}{\PGt\PGt}{}\HepParticle{\PH}{\PQb\PQb}{}}\xspace}
\newcommand{\zzbbtt}{\ensuremath{\HepParticle{\PZ}{\PQb\PQb}{}\HepParticle{\PZ}{\PGt\PGt}{}}\xspace}
\newcommand{\ZZZH}{\ensuremath{\ZZ/\ZH}\xspace}
\newcommand{\ZZbbtt}{\ensuremath{\PZ\PZ \to \PQb\PQb\PGt\PGt}\xspace}
\newcommand{\ZHbbtt}{\ensuremath{\PZ\PH \to \PQb\PQb\PGt\PGt}\xspace}
\newcommand{\ZZZHbbtt}{\ensuremath{\PZ\PZ / \PZ\PH \to \PQb\PQb\PGt\PGt}\xspace}
\newcommand{\XZZ}{\ensuremath{\HepParticle{X}{}{}\to\PZ\PZ}\xspace}
\newcommand{\XZZbbtt}{\ensuremath{\HepParticle{X}{}{}\to\zzbbtt}\xspace}
\newcommand{\ZprimeZH}{\ensuremath{\PZpr\to\PZ\PH}\xspace}
\newcommand{\Zprimezbbhtt}{\ensuremath{\PZpr\to\zbbhtt}\xspace}
\newcommand{\Zprimeztthbb}{\ensuremath{\PZpr\to\ztthbb}\xspace}

\newcommand{\HH}{\ensuremath{\PH\PH}\xspace}
\newcommand{\HHbbtt}{\ensuremath{\PH\PH\to \PQb\PQb\PGt\PGt}\xspace}

\newcommand{\FastMTT}{\textsc{FastMTT}\xspace}
\newcommand{\ParticleNet}{\textsc{ParticleNet}\xspace}

\newcommand{\fb}{\unit{fb}}
\newcommand{\pb}{\unit{pb}}

\newcommand{\resb}{\textit{resolved 1b}\xspace}
\newcommand{\resbb}{\textit{resolved 2b}\xspace}
\newcommand{\res}{\textit{resolved-1/2b}\xspace}
\newcommand{\boostedbb}{\textit{boosted-\bb}\xspace}

\newcommand{\boostedTauTau}{boosted \tautau}
\newcommand{\resolvedTaus}{resolved \tautau}
\newcommand{\catBoostedTaus}{\textit{boosted \bb{} - boosted \tautau}\xspace}
\newcommand{\catBoostedHPS}{\textit{boosted \bb{} - resolved \tautau}\xspace}

\newlength\cmsTabSkip\setlength{\cmsTabSkip}{1ex}

\cmsNoteHeader{B2G-25-004}
\title{Study of \ZZ and \ZH production in the \texorpdfstring{\bbtt}{b b-bar tau tau} final state and search for high-mass spin-0 and spin-1 resonances in proton-proton collisions at \texorpdfstring{$\sqrt{s}=13\TeV$}{sqrt(s) = 13 TeV}}

\date{\today}

\abstract{
A study of the production of pairs of \PZ bosons (\ZZ) and of the associated production of a \PZ boson and a Higgs boson (\ZH) in final states containing two \PQb quarks and two tau leptons (\bbtt) is presented. The analysis is based on proton-proton collisions collected at $\sqrt{s}=13\TeV$ by the CMS experiment at the LHC, corresponding to an integrated luminosity of 138\fbinv. The nonresonant analysis targets the standard model  \ZZ and \ZH processes in the \bbtt final state, motivated by the prominent role of this channel in searches for nonresonant Higgs boson pair production. The resonant searches target physics beyond the standard model, probing heavy spin-0 resonances \PX that decay into  \ZZ and spin-1 resonances \PZpr that decay into \ZH, with masses in the 0.2--5 and 0.5--6\TeV ranges, respectively. Upper limits at 95\% confidence level are set on the product of production rate and branching fraction $\sigma(\PX)\mathcal{B}(\PX\to\ZZ)$, ranging from 300\pb to 24\fb, and $\sigma(\PZpr)\mathcal{B}(\PZpr\to\ZH)$, ranging from 0.4\pb to 12\fb. These are the first measurements to probe the \ZZZHbbtt processes. No deviation from standard model expectations is observed.
}

\hypersetup{
pdfauthor={CMS Collaboration},
pdftitle={Study of ZZ and ZH production in the bbtautau final state and search for high-mass spin-0 and spin-1 resonances in proton-proton collisions at sqrt(s) = 13 TeV},
pdfsubject={CMS},
pdfkeywords={CMS, B2G, ZZ, ZH}}

\maketitle

\section{Introduction}\label{Sec:Intro}
The discovery of the Higgs boson (\PH) in 2012 by the ATLAS and CMS experiments~\cite{SingleHATLAS,SingleHCMS,SingleHCMSLongPaper} has provided a strong validation of the standard model (SM) of particle physics. However, critical aspects of the SM such as the self-coupling of the Higgs boson ($\lambda$) remain unmeasured to date. Higgs boson pair (\HH) production is directly sensitive to $\lambda$, which makes the measurement of \HH production a key goal of the high-luminosity program at the CERN LHC (HL-LHC)~\cite{NatureCMSHH,HiggsHLLHC}.
Among the various \HH decay channels, the final state with two \PQb quarks and two tau leptons (\bbtt) offers one of the highest sensitivities, motivating continued efforts to refine and strengthen the corresponding analysis strategies. In particular, studies of experimentally well established SM processes can help to improve the robustness and performance of \HHbbtt searches, especially as projections indicate that the full HL-LHC data set may enable the observation of \HH production~\cite{HiggsHLLHC,CMS:HHCombination,HLLHCprojections}.

The SM \ZZ and \ZH processes, which are experimentally well established through the leptonic \PZ boson decays~\cite{ZZCMSXS,ATLASZZRun2,ATLASZZ2022}, provide such benchmarks and can be studied in the \bbtt final state. The \ZZZHbbtt processes share the experimental challenges of the \HHbbtt search, while benefiting from significantly larger cross sections---approximately 64 times larger for \ZZ~\cite{ZZCMSXS} and 2.2 times larger for \ZH~\cite{CMS:ZHbb}.
A similar approach has been successfully applied in an $\PH\PH\to4\PQb$ search, where background modeling and analysis techniques were assessed using the $\ZZZH\to4\PQb$ processes~\cite{ZZZH4b}.

Beyond their use as standard candles, the \ZZZHbbtt processes can also be used to probe physics beyond the SM (BSM).
Several issues remain unaddressed by the SM, such as the baryon asymmetry of the universe and the various mass hierarchies, motivating extensions to the SM Lagrangian.
For example, BSM physics in the Higgs sector could manifest as one or more additional scalar fields, resulting in heavy resonances that decay into lower-mass bosons.
Examples of such BSM theories include the singlet model~\cite{Binoth:1996au,Schabinger:2005ei,Patt:2006fw}, the two-Higgs-doublet model~\cite{Branco:2011iw}, the heavy vector triplet model~\cite{HVTModel}, the minimal supersymmetric standard model~\cite{Fayet:1974pd,Fayet:1977yc}, and models with warped extra spatial dimensions~\cite{Agashe:2007zd,Fitzpatrick:2007qr}.

This paper presents measurements of various processes that lead to the \bbtt final state, using proton-proton collisions at $\sqrt{s} = 13\TeV$ collected by the CMS experiment during 2016--2018 and corresponding to an integrated luminosity of 138\fbinv.
Measurements using the \bbtt final state are presented for the SM production of \ZZ and for the search for a heavy scalar resonance that decays into \ZZ.
Measurements are also presented for the SM \ZH production and for the search for heavy spin-1 resonances that decay into \ZH, using the \bbtt final state.
Searches for heavy resonances that decay into \ZZ have been previously conducted in the fully leptonic~\cite{ATLAS:ResonantZZ4l,CMS:ResonantZZ4l}, semileptonic~\cite{ATLAS:ResonantZZllqqRun2,CMS:ResonantZZllqqRun2,CMS:ResonantZZqqnunuRun2}, and fully hadronic~\cite{ATLAS:ResonantZZHadronic,CMS:ZZZHResonancesHadronic} final states.
Searches for spin-1 resonances that decay into \ZH have been conducted in the $\PZ\to\PQq\PAQq, \PH\to\PGt\PGt$ channel using Lorentz-boosted topologies~\cite{CMS:ResonantZHToQQTauTauRun1,CMS:ResonantZHToQQTauTau2016}. Various other channels have also been studied, including $\PZ\to\PQq\PAQq, \PH\to\PQb\PAQb$~\cite{ATLAS:2020qiz,ATLAS:2023jdk,CMS:ZZZHResonancesHadronic} and $\PZ\to\Pell\Pell, \PH\to\PQb\PAQb$~\cite{ATLAS:2022enb,CMS:2024hbn} (where \Pell designates an electron or a muon).
This study is the first to probe \ZZ and \ZH decays to the \bbtt final state. It can probe possible enhancements in the coupling of a new \PZ boson-like particle to third-generation fermions~\cite{ZWithThirdGen}, providing an original contribution to these searches.
Other searches in the \bbtt final state have focused on spin-0 and spin-2 resonances that decay into \HH~\cite{ATLASRun2HHbbtt,ATLAS:ResonantHHboostedTaus,CMS:ResonantHHbbtautau2016,CMS:ResonantHHbbtautauBoostedTaus}.

Three decay modes of the \tautau system are considered: \muth, \eleth, and \thth, where \taumu, \taue, and \tauh correspond, respectively, to the decay of a tau lepton into a muon (\PGm), an electron (\Pe), or hadrons, with the associated neutrinos. The combination of these decay modes corresponds to 87.6\% of the possible \tautau decays.
Both resolved and boosted reconstruction regimes for the \tautau and \bb systems are considered.
The analysis strategy is similar to that of the \HHbbtt analysis~\cite{CMSRun2HHbbtt}.

Three different final states are considered:
\begin{itemize}
    \item \ZZbbtt (denoted \zzbbtt) with a decay branching fraction ($\mathcal{B}$) of 1.0\%~\cite{LHCHXSWG:Report4};
    \item $\PZ\PH \; (\PZ \to \PQb\PQb,\,\PH \to\PGt\PGt)$ (denoted \zbbhtt) with $\mathcal{B} = 0.9\%$~\cite{LHCHXSWG:Report4};
    \item $\PZ\PH \; (\PZ \to \PGt\PGt,\PH \to\PQb\PQb)$ (denoted \ztthbb) with $\mathcal{B} = 1.9\%$~\cite{LHCHXSWG:Report4}.
\end{itemize}

\begin{figure}[htbp]
    \centering
    \includegraphics[width=0.37\columnwidth]{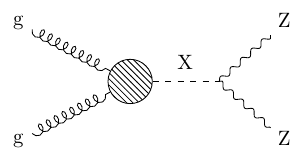}
    \includegraphics[width=0.37\columnwidth]{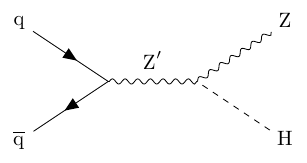}
    \caption{Illustrative diagrams describing the production of a high-mass resonance: a spin-0 resonance \XZZ (left) and a spin-1 resonance \ZprimeZH (right).}
    \label{fig:ResFeynman}
\end{figure}

Various hypotheses for the resonant searches are investigated, with heavy resonance masses ranging from 200\GeV to 5\TeV for spin-0 resonances, and from 500\GeV to 6\TeV for spin-1 resonances.
The narrow-width approximation is used to simplify the theoretical description and provide model-independent results. Illustrative diagrams are presented in Fig.~\ref{fig:ResFeynman}.

The paper is organized as follows. The CMS detector is described in Section~\ref{Sec:Detector} and the simulation of physics processes is summarized in Section~\ref{Sec:Simulation}, followed by a description of object reconstruction in Section~\ref{Sec:Reconstruction}. The event selection and categorization are described in Section~\ref{Sec:Selection}, and the approaches used to model backgrounds are described in Section~\ref{Sec:Background}. The systematic uncertainties are discussed in Section~\ref{Sec:Systematics}. Finally, the results are presented in Section~\ref{Sec:Results}, and the paper is summarized in Section~\ref{Sec:Summary}.
Tabulated results are provided in the HEPData record for this analysis~\cite{hepdata}.

\section{The CMS detector}\label{Sec:Detector}
The CMS apparatus~\cite{CMS:2008xjf,CMS:2023gfb} is a multipurpose, nearly hermetic detector, designed to trigger on~\cite{CMS:2020cmk,CMS:2016ngn,CMS:2024aqx} and identify electrons, muons, photons, and (charged and neutral) hadrons~\cite{CMS:2020uim,CMS:2018rym,CMS:2014pgm}. Its central feature is a superconducting solenoid of 6\unit{m} internal diameter, providing a magnetic field of 3.8\unit{T}. Within the solenoid volume are a silicon pixel and strip tracker, a lead tungstate crystal electromagnetic calorimeter (ECAL), and a brass and scintillator hadron calorimeter (HCAL), each composed of a barrel and two endcap sections. Forward calorimeters extend the pseudorapidity coverage provided by the barrel and endcap detectors. Muons are reconstructed using gas-ionization detectors interleaved with the layers of the steel flux-return yoke outside the solenoid. More detailed descriptions of the CMS detector, together with a definition of the coordinate system used and the relevant kinematic variables, can be found in Refs.~\cite{CMS:2008xjf,CMS:2023gfb}.

Events of interest are selected using a two-tiered trigger system.
The first level (L1), composed of custom hardware processors, uses information from the calorimeters and muon detectors to select events at a rate of around 100\unit{kHz} within a fixed latency of 4\mus~\cite{CMS:2020cmk}.
The second level, known as the high-level trigger, consists of a farm of processors running a version of the full event reconstruction software optimized for fast processing, and reduces the event rate to a few kHz before data storage~\cite{CMS:2016ngn,CMS:2024aqx}.

\section{Simulation of signal and background events}\label{Sec:Simulation}
{\tolerance=1200
Monte Carlo (MC) simulated samples are used to model both signals and most background processes.
The SM \ZZbbtt signal sample is generated at next-to-leading order (NLO) with \MGvATNLO v2.6.5~\cite{Alwall_2014}, and normalized to cross sections calculated at next-to-next-to-leading order (NNLO) in perturbative quantum chromodynamics (QCD)~\cite{ZZNNLOXS}. The two SM \ZH samples (\zbbhtt and \ztthbb) are generated separately at NLO using \POWHEG~\cite{powheg_1,powheg_2,powheg_3,powheg_zh} and normalized to cross sections calculated at NNLO in QCD~\cite{LHCHXSWG:Report4}.
Simulated samples of narrow-width spin-0 resonances \XZZbbtt are generated at NLO precision using \POWHEG~\cite{powheg_ggH}, with resonance masses ranging from 200\GeV to 5\TeV. Spin-1 resonances, \Zprimezbbhtt and \Zprimeztthbb, are generated with the heavy vector triplet model~\cite{HVTModel} using \MGvATNLO at leading order (LO), with resonance masses ranging from 500\GeV to 6\TeV.
\par}

{\tolerance=1200
The background composition depends on the analysis channels and categories, described in Section~\ref{Subsec:Categorization}, but is generally dominated by top quark pair production (\ttbar) and $\PZ/\PGgst \to \Pe\Pe/\PGt\PGt$ Drell--Yan (DY) production.
The \ttbar background is simulated at NLO with \POWHEG~\cite{powheg_ttbar}, whereas the DY background is simulated with \MGvATNLO at NLO with the FxFx jet merging~\cite{Frederix_2012}, considering up to two extra partons at the matrix element level.
Additional smaller background contributions are also considered.
The production of a \PW boson in association with jets (\PW{}+jets) is simulated with \MGvATNLO at LO precision with the MLM jet merging~\cite{Alwall_2007}, considering up to four extra partons at the matrix element level.
Single top quark (\PQt{}+\PX) and single Higgs boson (\PH{}+\PX) backgrounds are simulated at NLO precision with \POWHEG;
backgrounds from diboson ($\PW\PW$, \ZZ, and $\PW\PZ$) and triboson ($\PZ\PZ\PZ$, $\PZ\PZ\PW$, $\PZ\PW\PW$, and $\PW\PW\PW$) production are simulated at NLO with \MGvATNLO or \POWHEG;
background from Higgs boson pair production is simulated at LO with \MGvATNLO;
and production of \ttbar pairs in association with a boson ($\PQt\PAQt\PX$ where $\PX=\PW,\PZ,\PH$) is simulated with \MGvATNLO, either at LO or NLO depending on the process.
All background samples are normalized using the most precise cross section evaluations available, which are generally calculated at NLO or NNLO.
\par}

For both signal and background events, multiple parton interactions, parton shower, and hadronization effects, as well as the decays of the Higgs boson to either \PQb quarks or tau leptons for the signal, are simulated with \PYTHIA v8.230~\cite{Sj_strand_2015}, using the CP5~\cite{Sirunyan_2020} tune and the NNPDF 3.1~\cite{NNPDF_3_1} NNLO parton distribution functions (PDFs).
For all simulated samples, the CMS detector response is modeled with \GEANTfour~\cite{AGOSTINELLI2003250}, and additional collisions within the same or nearby bunch crossings (pileup) are simulated and superimposed on the hard-scattering process.
Events are then reweighted to match the pileup multiplicity distribution observed in data.

\section{Object reconstruction}\label{Sec:Reconstruction}
A particle-flow (PF) algorithm~\cite{ParticleFlow} aims to reconstruct and identify each individual particle in an event, with an optimized combination of information from the various elements of the CMS detector.
The energy of electron candidates is determined from the combination of the momentum at the primary interaction vertex as determined by the tracker, the energy of the corresponding ECAL cluster, and the energy sum of all bremsstrahlung photons spatially compatible with originating from the electron track. The momentum of muon candidates is obtained from the curvature of the corresponding track. The energy of charged hadrons is determined from the combination of the momentum measured in the tracker and the matching ECAL and HCAL energy deposits, corrected for the calorimeter response to hadronic showers. Finally, the energy of neutral hadrons is obtained from the corresponding corrected ECAL and HCAL energies.
The primary vertex (PV) is taken to be the vertex corresponding to the hardest scattering in the event, evaluated using tracking information alone, as described in Section~9.4.1 of Ref.~\cite{PrimaryVertexFromPhase2TDR}.

Muons are measured in the pseudorapidity range $\abs{\eta}<2.4$, with detection planes made using three technologies: drift tubes, cathode strip chambers, and resistive plate chambers. The \pt resolution of muons ranges from 1 to 7\% depending on the pseudorapidity and \pt~\cite{CMS:2018rym}.
Muon candidates are identified based on the number of hits in the tracker and in the muon systems, and on the compatibility of the track reconstructed in the silicon tracker with the hits in the muon detector and with originating from the PV.

Electrons are reconstructed in the $\abs{\eta}<2.5$ range, and their momentum is estimated by combining the energy measurement in the ECAL with the momentum measurement in the tracker.
The momentum resolution for electrons with $\pt \approx45\GeV$ from $\PZ\to\Pe\Pe$ decays ranges from 1.6 to 5.0\%~\cite{CMS:2020uim,CMS-DP-2020-021}.
Electron identification (ID) is based on a combination of purely-ECAL, purely-tracker, and ECAL-plus-tracker observables.

Each reconstructed electron or muon is required to have a distance between the track of the candidate and the PV in the transverse and longitudinal planes satisfying $\abs{d_{xy}}<0.045\unit{cm}$ and $\abs{d_z}<0.2\unit{cm}$, respectively.
Electrons and muons are also required to satisfy isolation requirements, which depend on the kinematic regime of the \tautau pair, and are described below. They are based on the PF relative isolation ($I_{\text{rel}}^\Pell$)~\cite{ParticleFlow}, computed by summing the transverse momenta of PF candidates in a cone around the lepton direction. The cone radius is $\Delta R = 0.4\,(0.3)$ for muons (electrons), where $\Delta R = \sqrt{\smash[b]{\Delta\eta\,^2 + \Delta\phi\,^2}}$. The isolation is corrected for pileup contributions as shown below, where $\mathrm{h}_{\mathrm{PV}}$ designates hadrons originating from the hard scatter, and $\mathrm{h}^\pm_{\mathrm{PU}}$ designates charged hadrons from pileup vertices:
\begin{equation}
    I_{\text{rel}}^\Pell = \left(\sum \pt^{\mathrm{h}_{\mathrm{PV}}} + \sum \pt^\PGg - 0.5\sum \pt^{\mathrm{h}^\pm_{\mathrm{PU}}} \right) / \pt^\Pell
\end{equation}

In the resolved kinematic regime of the \tautau pair (corresponding to the \resolvedTaus{} categories defined in Section~\ref{Subsec:Categorization}), where the lepton is expected to be isolated, standard relative isolation criteria are applied.
For muons, candidates must satisfy $I_{\text{rel}}^{\PGm}<0.15$ and the tight ID criteria described in Ref.~\cite{CMS:2018rym}.
For electrons, a boosted decision tree (BDT) is employed for ID, which includes isolation variables. The ``tight'' working point of this discriminator, characterized by an 80\% efficiency, is employed~\cite{CMS:2020uim}.

In the boosted kinematic regime of the \eleth (\muth) channel, denoted \boostedTauTau{}, the electron (muon) and \tauh are expected to be close to each other. Thus, dedicated isolation criteria are applied, where the corrected relative isolation ($I_{\text{corr}}^{\Pell}$) is computed excluding components from the adjacent \tauh from the sum. Muons are required to satisfy the loose ID criteria~\cite{CMS:2018rym} and to satisfy $I_{\text{corr}}^{\PGm}<0.25$.
For electrons, the loose working point of the cut-based ID~\cite{CMS:2020uim} is used, corresponding to a nominal efficiency for electrons from $\PZ\to\Pe\Pe$ decays of 90\%. The isolation requirement of the cut-based ID is replaced with a dedicated isolation selection requiring $I_{\text{corr}}^{\Pe} < 0.112 + (0.506\GeV)/\pt$ in the barrel region ($\abs{\eta}<1.479$) and $I_{\text{corr}}^{\Pe} < 0.108 + (0.963\GeV)/\pt$ in the endcap region ($1.479<\abs{\eta}< 2.5$).

Jets are reconstructed from PF candidates, clustered using the anti-\kt algorithm~\cite{Cacciari:2008gp,Cacciari:2011ma}. Two values of the distance parameter $R$ are used, 0.4 for resolved topologies and 0.8 for boosted topologies. These yield the ``AK4'' and ``AK8'' jet collections, respectively.
Pileup interactions can contribute extra tracks and calorimetric energy depositions, increasing the apparent jet momentum.
To mitigate this effect in AK4 jets, tracks identified to originate from pileup vertices are discarded and an offset correction is applied to correct for remaining contributions~\cite{Sirunyan:2020foa}.
For AK8 jets, pileup mitigation is carried out through the pileup-per-particle ID algorithm~\cite{Bertolini:2014bba}, acting at the reconstructed particle level and making use of local shape information, event pileup properties, and tracking information.
The jet momentum is determined as the vectorial sum of all particle momenta in the jet.
Jet energy corrections are derived so that the average measured energy of jets becomes identical to that of particle-level jets, and to correct residual differences in jet energy scale between data and simulation~\cite{Khachatryan:2016kdb,CMS-DP-2021-033}.
The jet energy resolution typically amounts to 15--20\% at 30\GeV, 10\% at 100\GeV, and 5\% at 1\TeV~\cite{Khachatryan:2016kdb}.
Additional selection criteria are applied to each jet to remove jets potentially dominated by instrumental effects or reconstruction failures~\cite{Sirunyan:2020foa}.

The missing transverse momentum vector \ptvecmiss is computed as the negative vector sum of the transverse momenta of all the PF candidates in an event, and its magnitude is denoted as \ptmiss~\cite{METPerf}. The \ptvecmiss is modified to account for corrections to the energy scale of the reconstructed jets in the event, and is used to infer the momentum of neutrinos originating from tau lepton decays.

The AK4 jets originating from \PQb quarks are identified using the \textsc{DeepJet} algorithm~\cite{Bols_2020,BTVDeepJetDPS}, a deep neural network combining secondary vertex properties, track-based variables, and PF jet constituents (neutral- and charged-particle candidates).
The medium working point of the \textsc{DeepJet} \PQb-tagging discriminant is used, characterized by a 73--80\% \PQb jet efficiency depending on data-taking period, and a 1\% light jet mistagging efficiency.
The distribution of the discriminant in simulation is corrected to match the distribution measured in data~\cite{BTV-16-002}.
For the identification of AK8 jets originating from the decay of a high transverse momentum \PZ or Higgs boson into a pair of \PQb quarks, the \ParticleNet algorithm~\cite{ParticleNet} is used. This deep learning approach represents jets as unordered sets of constituent particles, or ``particle clouds'', and makes use of a dynamic graph convolutional neural network to exploit their spatial and kinematic correlations. A threshold on the \ParticleNet output score is applied to select jets originating from a resonance decaying into a pair of \PQb quarks and reject background from QCD multijet production. The low-purity working point is used, corresponding to a signal jet efficiency of 80\%~\cite{BoostedBBJetIDPerf}.
Corrections are applied to simulation to match the efficiency measured in data. For genuine \PQb jets from resonant decays (\eg, a Higgs boson decaying into a pair of \PQb quarks), corrections are derived from gluon-splitting $\Pg\to\PQb\PAQb$ proxy jets and from Lorentz-boosted \PZ decays into $\PQb\PAQb$, as described in Ref.~\cite{BoostedBBJetIDPerf}. For jets originating from other topologies, such as from DY or \ttbar processes, channel-specific corrections are computed in control regions enriched in those processes.

The constituents of the AK8 jets are reclustered using the Cambridge--Aachen algorithm~\cite{Dokshitzer:1997in,Wobisch:1998wt}; the soft-drop algorithm~\cite{Dasgupta:2013ihk,Butterworth:2008iy} is applied to remove soft, wide-angle radiation from the jet. Signal AK8 jets are required to have a groomed jet mass ${>} 30\GeV$ in order to further suppress QCD multijet background.

Hadronic decays of tau leptons are reconstructed from jets, using the ``hadrons-plus-strips'' (HPS) algorithm~\cite{TauPerfRun2}, which combines one or three tracks with energy deposits in the calorimeters to identify the tau decay modes. Neutral pions are reconstructed as strips with dynamic size in $\eta$-$\phi$ from reconstructed electrons and photons, where the strip size varies as a function of the \pt of the electron or photon candidate. The strips are narrow in $\eta$ but wide in $\phi$.
The spread in $\phi$ accounts for photons originating from neutral pion decays that convert into electron-positron pairs while traversing the silicon tracker.
Their trajectories are bent in opposite directions in $\phi$ by the magnetic field and they may emit further bremsstrahlung photons before reaching the ECAL.
The decay modes considered in this analysis produce one charged pion or kaon together with zero, one, or two neutral pions; or three charged pions or kaons together with zero or one neutral pion. The tau candidates are required to have a distance between the track of the candidate and the PV in the longitudinal plane satisfying $\abs{d_z}<0.2\unit{cm}$.

When a \tautau pair is highly collimated, such as when it results from the decay of a high \pt \PZ or Higgs boson, the reconstruction becomes challenging, necessitating dedicated boosted reconstruction algorithms~\cite{TauPerfRun2}.
The procedure starts from a large-radius jet clustered with the Cambridge--Aachen algorithm~\cite{Wobisch:1998wt} with a distance parameter of 0.8, and $\pt>100\GeV$.
The last step of the clustering is reversed to form two subjets. If they both satisfy $\pt>10\GeV$ and the mass drop condition $\max(m_{\text{subjet},1}, m_{\text{subjet},2})/m_{\text{CA8 jet}} < 2/3$, the two subjets are passed to the HPS algorithm as potential seeds for \tauh reconstruction.
In this ratio, $m_{\text{subjet},1}$ and $m_{\text{subjet},2}$ designate the masses of the two subjets, while $m_{\text{CA8 jet}}$ designates the mass of the original CA8 jet.
If the \pt and mass drop conditions are not met, the procedure is restarted in an iterative fashion, considering the most energetic subjet as a starting point.
Semileptonic \eleth and \muth pairs can also be reconstructed in this way, where the electron or muon forms one subjet and the \tauh forms the other subjet. The \tauh subjet is then passed to the HPS algorithm as a seed.
This method is referred to as \boostedTauTau reconstruction, and was previously used by the CMS Collaboration in analyses involving $\PH\to\tautau$ decays~\cite{CMS:ResonantZHToQQTauTauRun1,CMS:SMHiggsTauTauBoosted,CMS:ResonantHHbbtautauBoostedTaus}.

To distinguish genuine \tauh decays from jets originating from the hadronization of quarks or gluons, the \textsc{DeepTau} algorithm~\cite{CMS:2022prd} is used.
Information from all individual reconstructed particles (hadrons, electrons, photons, muons) near the \tauh axis ($\Delta R<0.5$) is combined with properties of the \tauh candidate and of the event, and fed into a convolutional neural network.
For resolved \tauh, three different discriminants are employed in order to distinguish \tauh candidates from jets (\textsc{DeepTauVsJets}), electrons (\textsc{DeepTauVsEle}), and muons (\textsc{DeepTauVsMu}). For boosted \tauh candidates, a dedicated training of \textsc{DeepTau} targeting boosted topologies, known as \textsc{BoostedDeepTau}~\cite{DeepBoostedTauDPS}, is used. The \textsc{BoostedDeepTauVsJets} discriminant is used for rejection of jets, and the \tauh are additionally required not to overlap with a loosely reconstructed muon or electron.

\section{Event selection and categorization}\label{Sec:Selection}
\subsection{Trigger selection}\label{Subsec:Triggers}
The analysis considers three decay modes of the \tautau system (\muth, \eleth, and \thth) under two reconstruction regimes: \resolvedTaus and \boostedTauTau.
In the \resolvedTaus regime, different trigger strategies are employed for each channel. Single-lepton and lepton-\tauh cross triggers are used for the \muth and \eleth channels. The former requires the presence of only an isolated electron or muon, whereas the latter also requires an additional hadronically decaying tau lepton. Using cross triggers improves the analysis sensitivity by allowing the \pt threshold on electrons and muons to be lowered. In the \thth channel, the trigger requires the presence of a pair of hadronically decaying tau leptons with $\pt>35\GeV$.
The high-level trigger thresholds for electrons and muons are in the 24--32 and 19--27\GeV ranges, respectively, depending on the data-taking period. The lower bound of the range comes from cross triggers. Trigger thresholds on the \pt of the \tauh for the cross triggers are in the 20--30\GeV range.
Compared to the online selection, the offline thresholds are increased by 2\GeV for muons, 1\GeV for electrons, and 5\GeV for \tauh candidates. These values are chosen to ensure that the trigger efficiencies reach their plateau values.

At high resonance masses, the \tautau decay products become collimated, and triggers requiring isolation become inefficient. A \ptmiss trigger is therefore used for the \boostedTauTau regime for all channels. The online \ptmiss threshold is 110 or 120\GeV depending on the data-taking period, and the offline threshold on \ptmiss is 180\GeV.

Because of the limited time resolution of the various subdetectors involved, objects at the L1 trigger can be assigned to the wrong bunch crossing, which can lead to a loss of events (``trigger prefiring''). While normally negligible, a slowly increasing shift of the reconstructed cluster time in the ECAL occurred in the L1 trigger during the 2016 and 2017 data-taking periods, predominantly at high $\abs{\eta}$. The resulting efficiency loss is not included in simulation, and simulated events are reweighted to account for this inefficiency~\cite{CMS:2024ppo}. Similarly, some trigger prefiring affected the L1 muon trigger during 2016 and, to a lesser extent, during 2017--2018 data-taking, and it is accounted for in the analysis.

\subsection{Categorization}\label{Subsec:Categorization}
Events are classified in three different \tautau decay channels based on the information from the \PGt candidates reconstructed offline.
Selected signal events are required to have at least one \tauh candidate.
An event is classified as \muth if a muon satisfying the selections detailed in Table~\ref{tab:selections} is found; otherwise it is classified as \eleth if an electron satisfying the selections is found.
If neither a muon nor an electron is found, the event is classified as \thth if two \tauh candidates are found.
To resolve possible ambiguities, pairs are sorted according to isolation, favoring the most isolated muon (electron) in the \muth (\eleth) case or the most isolated \tauh in the \thth case.
After the three \resolvedTaus{} channels have been tried, the three channels are attempted again in the same order with \boostedTauTau reconstruction.

The two objects forming a pair, referred to as the ditau pair in the following, must have opposite charge and satisfy $\Delta R$ requirements depending on the channel and reconstruction regime. To reduce the $\PZ/\PGg^\ast\to\Pe\Pe/\PGm\PGm$ contamination, events with additional electrons or muons are discarded. The selection requirements on the additional leptons are looser than the nominal selections.
\begin{table*}[htbp]
    \centering
    \topcaption{Summary of selections applied to the \tautau candidate pair}
    \renewcommand{\arraystretch}{1.08}
    \begin{tabular}{p{3.4cm}p{6.1cm}p{5.2cm}}
        & Resolved \tautau & Boosted \tautau \\
        \hline
        Muons &$\pt > 21\text{--}29\GeV$, $\abs{\eta}<2.1$ & $\pt > 20\GeV$, $\abs{\eta}<2.4$ \\
        Electrons & $\pt > 25\text{--}33\GeV$, $\abs{\eta}<2.5\,(2.1)$\newline for single (cross) trigger & $\pt > 20\GeV$, $\abs{\eta}<2.5$ \\
        Hadronic taus &$\pt > 25\text{--}40\,(20) \GeV$, $\abs{\eta}<2.1\,(2.3)$\newline for ditau and cross (single) triggers & $\pt > 40\GeV$, $\abs{\eta}<2.3$ \\[\cmsTabSkip]
        Muon ID        & tight muon ID & loose muon ID \\
        Electron ID    & tight electron BDT ID & loose electron cut-based ID\newline (excluding isolation selection) \\
        \tauh ID\newline (\muth channel) & medium \textsc{DeepTauVsJet}\newline tight \textsc{DeepTauVsMu}\newline very-very-loose \textsc{DeepTauVsEle} & \textsc{BoostedDeepTauVsJet}\\[\cmsTabSkip]
        \tauh ID\newline (\eleth channel) & medium \textsc{DeepTauVsJet}\newline tight \textsc{DeepTauVsMu} \newline tight \textsc{DeepTauVsEle} & \textsc{BoostedDeepTauVsJet}\\[\cmsTabSkip]
        \tauh ID\newline (\thth channel)         & medium \textsc{DeepTauVsJet}\newline very-loose \textsc{DeepTauVsMu}\newline very-very-loose \textsc{DeepTauVsEle} & \textsc{BoostedDeepTauVsJet}\\[\cmsTabSkip]
        Pair requirement                  & opposite-charge \PGt candidates & opposite-charge \PGt candidates \\
        Separation between \tautau pair candidates  & $\Delta R>0.5$ & $0.05<\Delta R<0.7$ (\muth only)\newline $0.05<\Delta R<0.6$ (\eleth only)\newline $0.05<\Delta R<0.8$ (\thth only) \\
    \end{tabular}
    \label{tab:selections}
\end{table*}

The \ZHbb candidates are then identified by requiring selected events to have at least two additional AK4 jets with $\pt>20\GeV$, $\abs{\eta}<2.4$, and $\Delta R(j,\Pell)>0.5$, where \Pell denotes each of the two leptons forming the ditau pair. The \textsc{HH-bTag} algorithm~\cite{CMSRun2HHbbtt} is used to select the pair of \PQb jets originating from the Higgs or \PZ boson decay. The algorithm was developed for the \HHbbtt analysis and is based on deep neural networks (DNNs). The input features include the kinematic properties of the AK4 jets and of the \ZHtt system, the \textsc{DeepJet} scores of the \PQb jet candidates, and their angular separation with respect to the \ZHtt candidate. For each event, a score is assigned by the \textsc{HH-bTag} algorithm for all \PQb jet candidates, and the two jets with the highest scores are taken as those originating from the decay of the \PZ or Higgs boson.

To increase the sensitivity to boosted \PZ/Higgs bosons produced in the decay of heavy resonances and that decay into \bb, events containing one AK8 jet with $\pt>250\GeV$, $\abs{\eta}<2.4$, and $\Delta R(j,\Pell)>0.8$ are also retained. The \ParticleNet algorithm~\cite{ParticleNet} is used to discriminate jets compatible with a \ZHbb decay from those originating from multijet production.

Resolved \tautau events are further classified into three mutually exclusive categories, exploiting the large range of Lorentz boosts of the \ZHbb system. Events with two AK4 \PQb jet candidates passing the medium \textsc{DeepJet} working point are assigned to the \resbb category. Among events failing the \resbb selection, those containing an AK8 jet passing the low-purity \ParticleNet working point define the \catBoostedHPS category. Events failing both the \resbb and \boostedbb requirements, but in which one of the \PQb jet candidates passes the medium \textsc{DeepJet} working point, are assigned to the \resb category. Events in which no \PQb jet is identified are discarded, resulting in a negligible loss of signal while significantly reducing the multijet background contribution.
Events classified as \boostedTauTau are all required to contain an AK8 jet, defining the \catBoostedTaus category.

This defines a total of four reconstruction categories. Together with the three \tautau decay channels, this results in 12 mutually exclusive categories.

\subsection{\ZHtt mass reconstruction and invariant mass selections}\label{Subsec:mass_sels}
The invariant mass of the \tautau system, $m_{\PGt\PGt}$, is a key variable to separate \PZ/\PH candidate events from the background in this analysis. The \FastMTT algorithm~\cite{FastMTT} is used to reconstruct the four-momentum of the \tautau system. It uses \ptvecmiss and its resolution, as well as the four-vectors of the reconstructed visible \PGt decay products, to improve the resolution in $m_{\PGt\PGt}$. To reduce the computation time, it applies the collinear approximation~\cite{CollinearApprox}, which assumes that the neutrinos from \PGt decays are collinear with the visible decay products.

\begin{figure*}[htbp]
    \centering
    \includegraphics[width=0.45\textwidth]{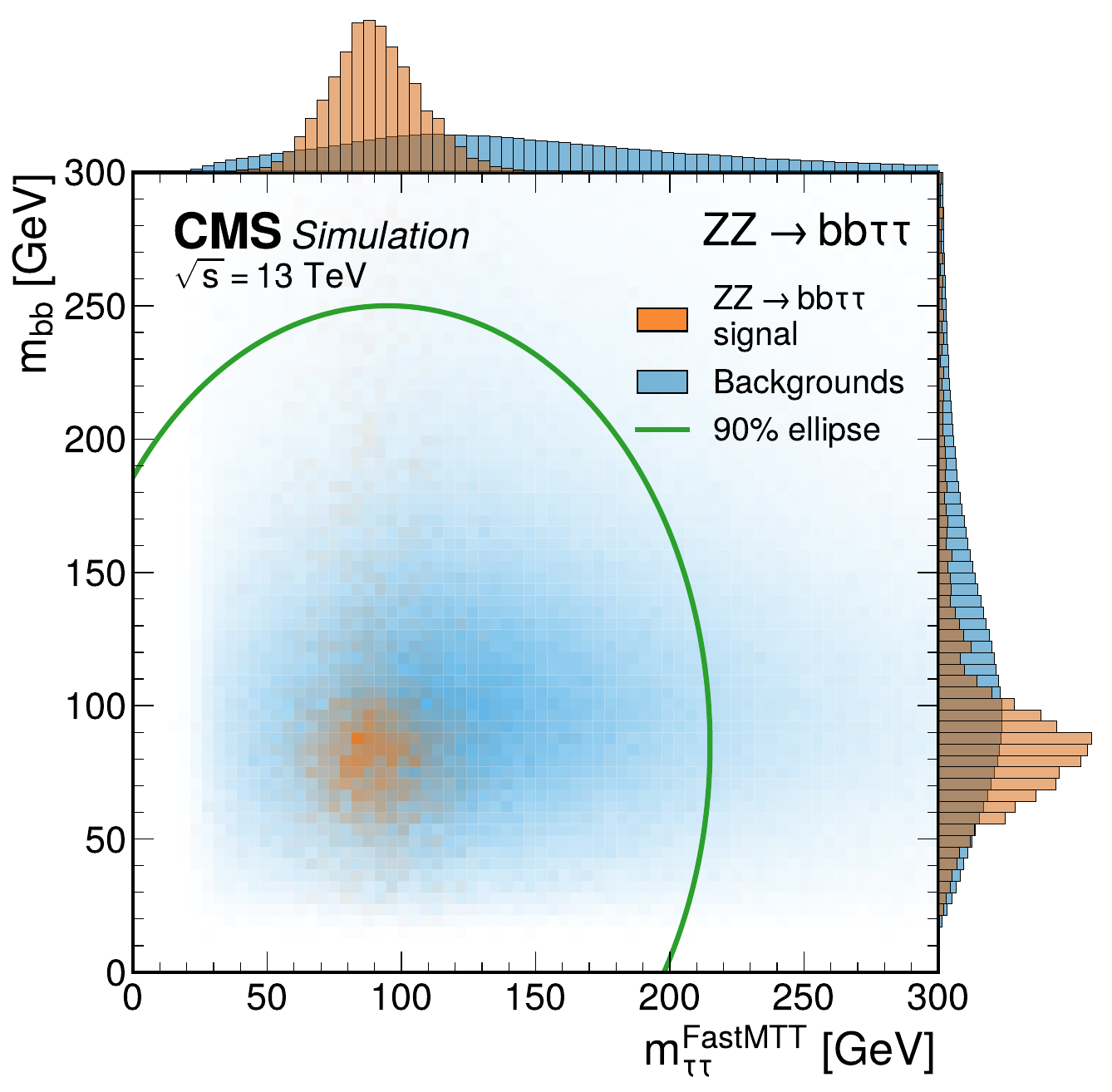}
    \includegraphics[width=0.45\textwidth]{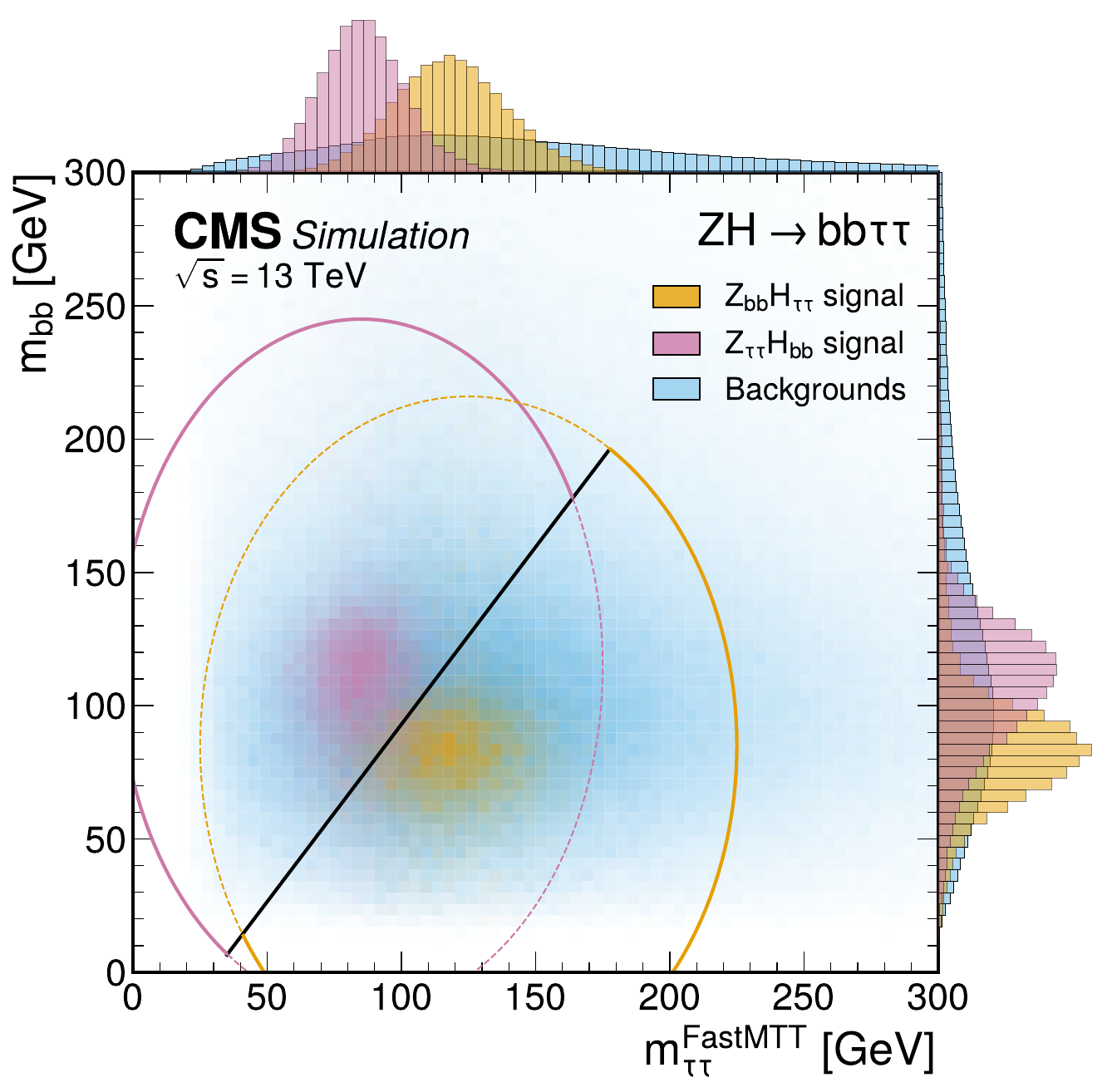}
    \caption{Distributions in the ($m_{\PQb\PQb},m_{\tautau}^{\FastMTT}$) plane for the SM \ZZ (left) and \ZH (right) signals, and for the sum of all simulated backgrounds (blue). The corresponding one-dimensional projections are also shown.
    The distributions are normalized to unity after the \tautau and $\PQb\PQb$ candidate selections have been applied.
    The colored curves indicate the elliptical mass selection, and the black line in the right panel marks the boundary between the regions targeting each of the two \ZH processes.}
    \label{fig:EllipticalCut}
\end{figure*}

The distribution of $m_{\PQb\PQb}$ with respect to $m_{\tautau}^{\FastMTT}$ is shown in Fig.~\ref{fig:EllipticalCut}. The quantity $m_{\PQb\PQb}$ is computed from the dijet invariant mass in the \res categories and from the soft-drop groomed mass~\cite{Larkoski:2014wba} in the \boostedbb categories.

An elliptical selection in the $m_{\PQb\PQb}$-$m_{\tautau}^{\FastMTT}$ plane is applied, as shown in Fig.~\ref{fig:EllipticalCut}, with a signal efficiency ${>}90\%$. Additionally, for the \ZH analysis, a linear selection (black line in the right panel of Fig.~\ref{fig:EllipticalCut}) is used to define two orthogonal parameter space regions targeting the \zbbhtt and \ztthbb processes, respectively. The two processes are combined when extracting the final results.

\section{Background modeling}\label{Sec:Background}
The analysis searches for final states characterized by the presence of two \PQb quarks and two tau leptons. The primary background contributions arise from \ttbar production, DY processes (mostly $\PZ/\PGgst \to \Pe\Pe$ and $\PZ/\PGgst \to \tautau$), and QCD multijet events (denoted QCD background). Additional background contamination originates from single top quark production (denoted \PQt, \PQt{}+\PW), \PW boson production in association with jets (denoted \PW{}+jets), as well as diboson and triboson (denoted $\PV\PV(\PV)$), \ttbar production in association with a single boson or a pair of vector bosons (denoted $\PQt\PQt\PX$), Higgs boson production in association with a vector boson (denoted $\PV\PH$), and electroweak production of a \PZ boson and two jets (denoted EWK \PZ).
Backgrounds are estimated using simulated samples, with the exception of the QCD multijet background, which is estimated using data.

The QCD background is estimated from data using the ABCD method, which involves defining four distinct regions within the analysis phase space.
The signal region (A) is defined by the signal selection criteria. The three control regions (B, C, and D) are enriched in QCD background, and are built as follows.
Control region B is built by requiring the two objects forming the ditau pair to have the same electric charge.
Control region C is constructed by inverting the selection on the \tauh discriminator against jets (\textsc{DeepTau} in the \resolvedTaus categories, or \textsc{BoostedDeepTau} in the \boostedTauTau category). In the \thth channel, only the lower-\pt \tauh candidate has its selection inverted.
Control region D is constructed by requiring a same-charge ditau pair and inverting the \tauh discriminator against jets.

The key assumption of this method is that the ratio of events in regions C and D can be used to predict the background contamination in region A from the number of events in the background-dominated region B. This provides an estimate of the QCD background in the signal region as $\mathrm{B}\, \mathrm{C}/\mathrm{D}$. The shape of the QCD background is extracted from region B, whereas the normalization is determined using the yields in regions B, C, and D. In the B, C, and D regions, the backgrounds estimated from simulation are subtracted from the observed data yields before the method is applied.

\begin{figure}[htbp]
    \centering
    \includegraphics[width=0.5\textwidth]{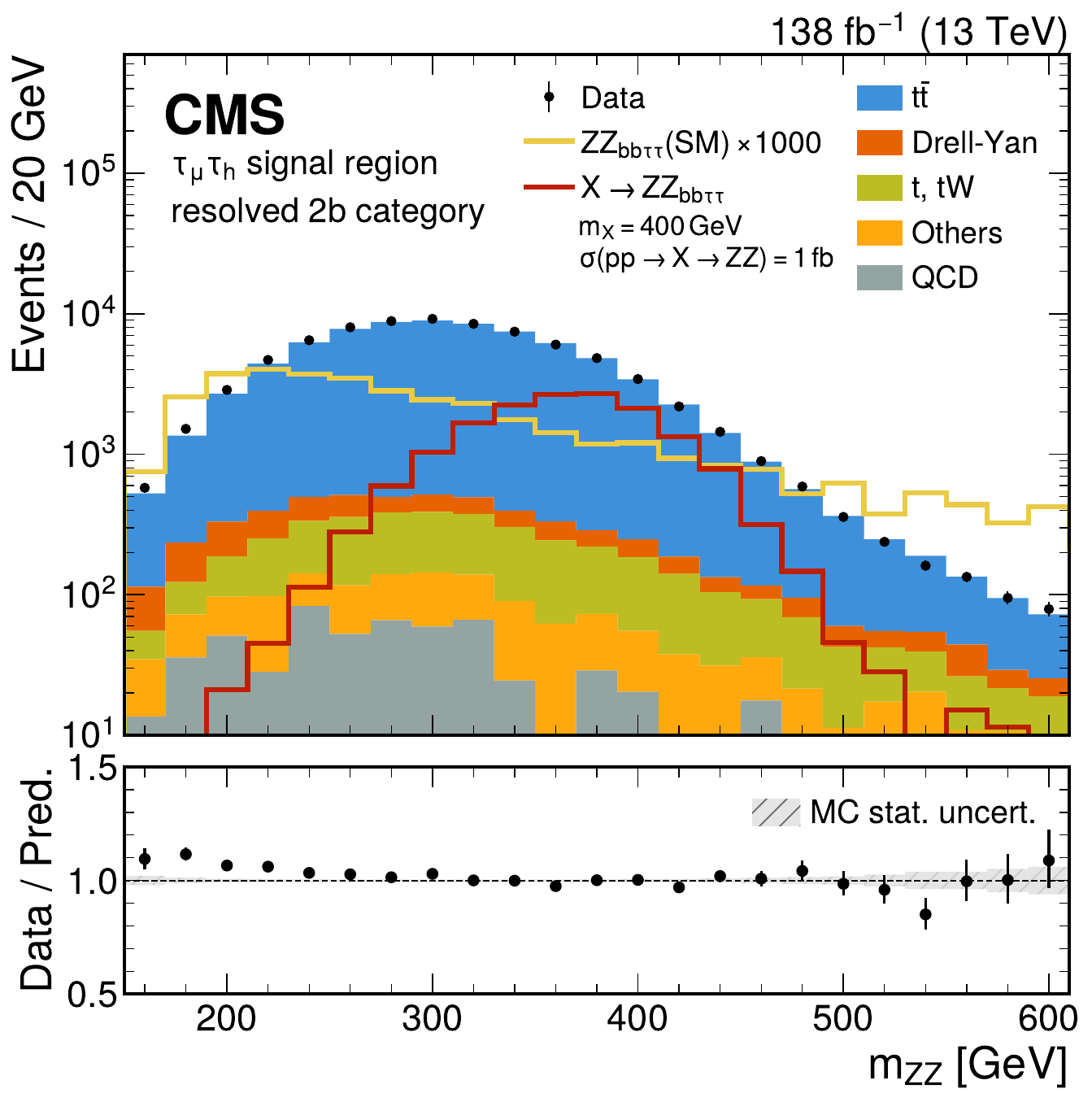}
    \caption{Distribution of the invariant mass of the \ZZ system, in the \resbb category of the \muth channel. The four-momentum of the \tautau system is reconstructed using \FastMTT. The SM \ZZbbtt signal, as well as the resonant \XZZbbtt signal for $m_\PX=400\GeV$, are overlaid, scaled for visibility. The lower panel shows the ratio of the data to the total SM prediction.}
    \label{fig:MZZ}
\end{figure}

The approaches considered for modeling the various background contributions yield good agreement with data. Figure~\ref{fig:MZZ} shows the invariant mass of the \ZZ system in one of the signal regions.

\section{Final discriminant}\label{Subsec:Discriminant}
Different DNNs are trained for signal extraction. For each of the SM processes \zzbbtt, \zbbhtt, and \ztthbb, a dedicated network is trained to target the corresponding signal process. For the resonant signals \XZZbbtt, \Zprimezbbhtt, and \Zprimeztthbb, a parameterized neural network~\cite{parameterised_nn} is trained for each signal type, using all generated mass points for training and the resonance mass as an input parameter to the network.
The training uses the signal and background samples described in Section~\ref{Sec:Simulation}, as well as all \tautau decay channels and jet categories.

The model is trained using both categorical and continuous input features, for a total of 27 features.

\textit{Categorical features:}
\begin{itemize}
\item \tautau decay mode (\muth, \eleth, or \thth) and \ZHtt reconstruction mode (resolved or boosted);
\item \ZHbb reconstruction mode (resolved or boosted);
\item data-taking period.
\end{itemize}
\textit{Continuous features:}
\begin{itemize}
\item \PQb tagging scores of the AK4 \PQb jet candidates;
\item invariant masses and transverse momenta of the \ZZZH, \tautau, and \bb systems;
\item \DR separation between the two tau lepton candidates;
\item kinematical variables, such as visible momenta and distances in the $\eta$-$\phi$ plane between reconstructed particles, describing the spatial distribution of the physics objects in the event;
\item resonance mass (for the resonant search only, parameterized input feature).
\end{itemize}

The DNN architecture~\cite{perceptron,Linnainmaa_70,Werbos_81,backprop} is composed of two discriminators, each consisting of ten neural networks trained using ten-fold stratified cross-validation of the training set.
For training, the simulated events are split evenly into two subsets, and each discriminator is trained on a different subset. At inference, each discriminator is applied to the subset of events not used in its training. As a result, all simulated events contribute to the final template distributions, reducing the associated statistical uncertainties, while avoiding biases that would otherwise arise if a discriminator were applied to events used in its own training.

\begin{figure}[htbp]
	\centering
	\includegraphics[width=0.5\textwidth]{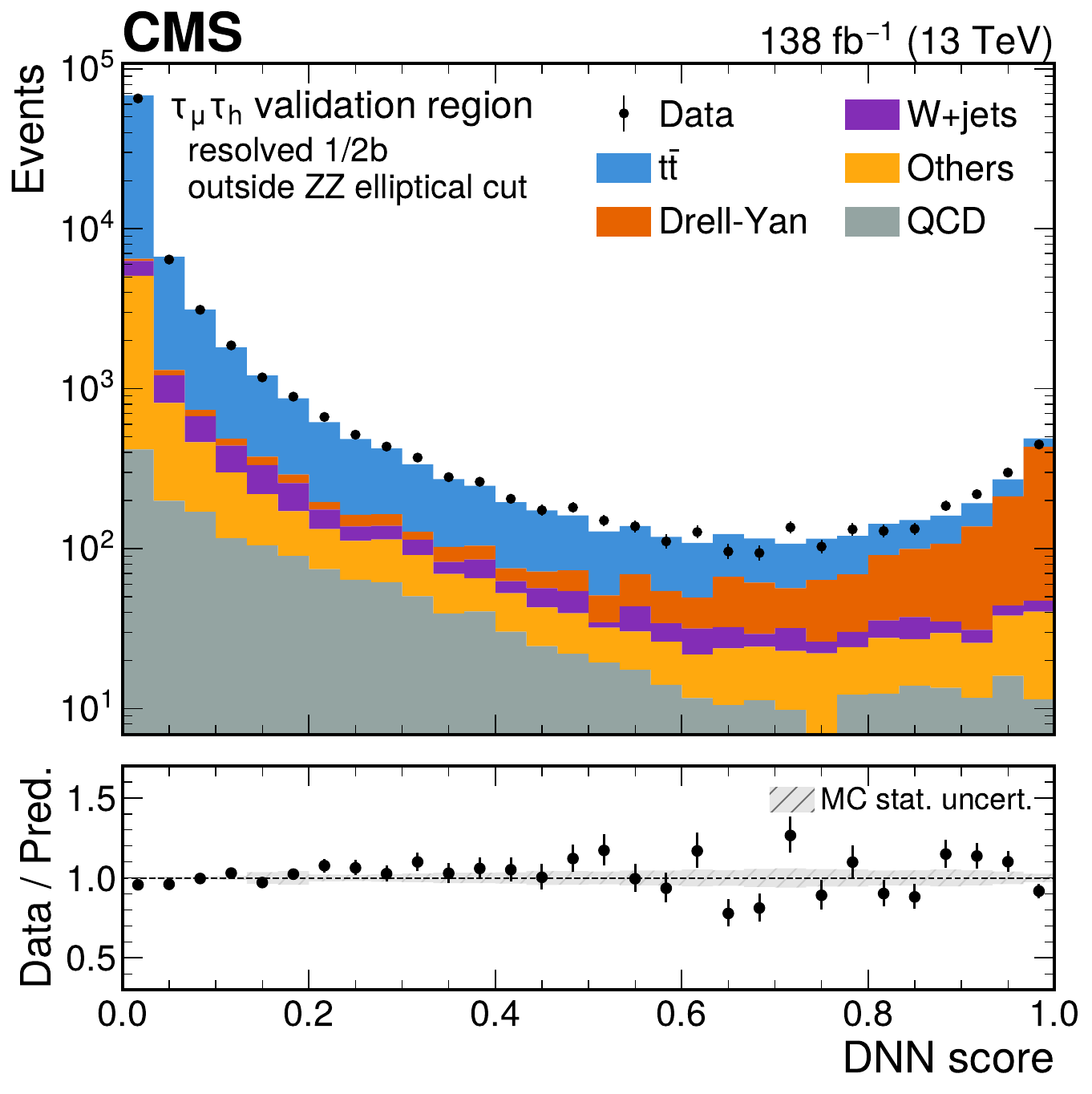}
	\caption{Distribution of the nonresonant DNN used for \ZZbbtt signal extraction in a validation region defined by inverting the elliptical mass selections described in Section~\ref{Subsec:mass_sels}, for the \res categories in the \muth channel. The lower panel shows the ratio of the data to the total SM prediction.}
	\label{fig:DNNdistributionsCRZZ}
\end{figure}

Figure~\ref{fig:DNNdistributionsCRZZ} shows the agreement between data and simulation of the DNN distribution in a validation region built by inverting the elliptical mass selection defined in Section~\ref{Subsec:mass_sels}.

\begin{figure}[htbp]
	\centering
	\includegraphics[width=0.7\textwidth]{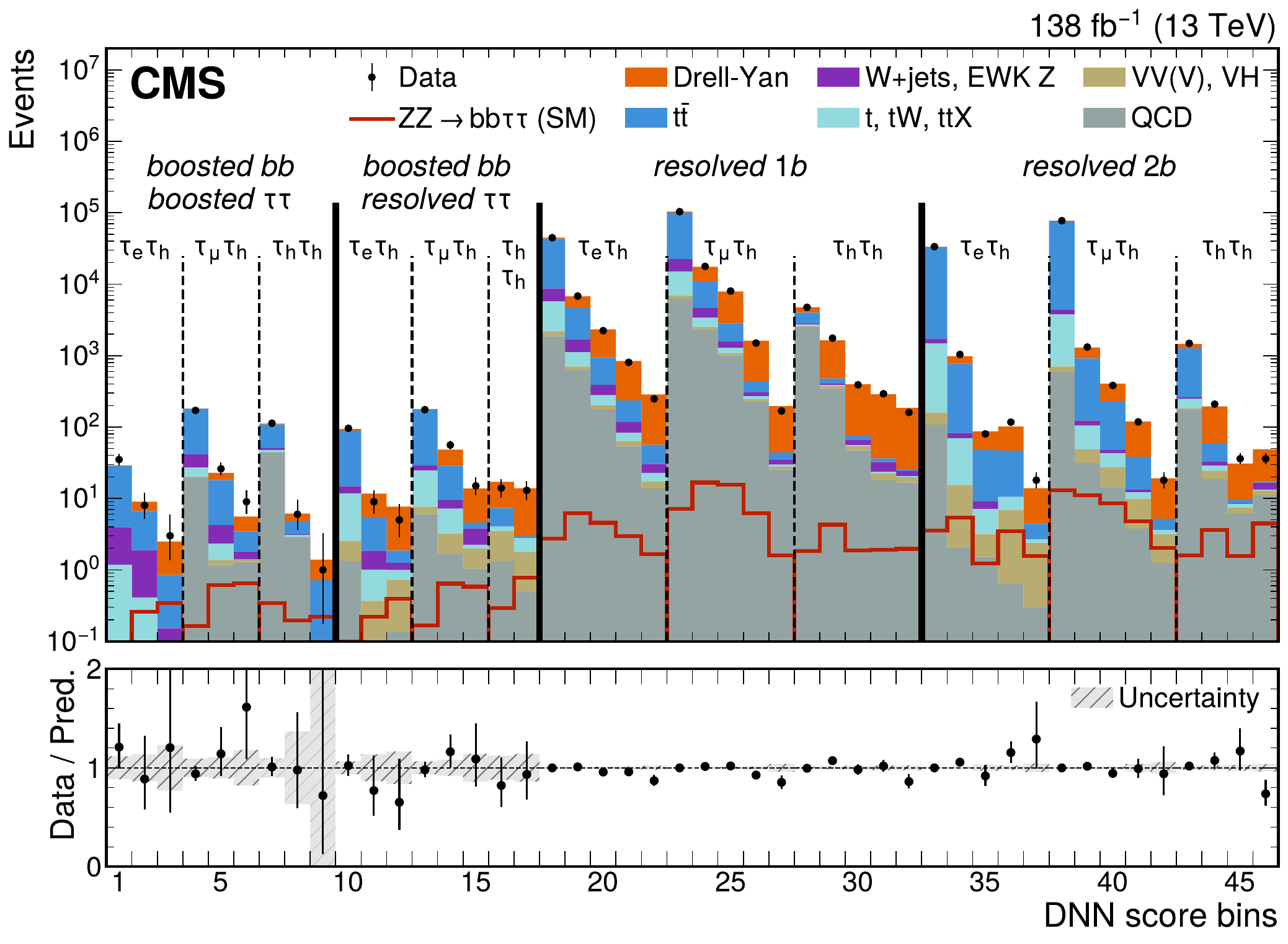}
	\caption{Distributions of the output of the DNN used for SM \ZZbbtt signal extraction. The lower panel shows the ratio of the data to the background prediction with uncertainties from a background-only fit to the observed data. The \ZZ signal is scaled to its SM cross section.}
	\label{fig:DNNdistributionsNonResZZ}
\end{figure}

\begin{figure}[htbp]
	\centering
	\includegraphics[width=0.75\textwidth]{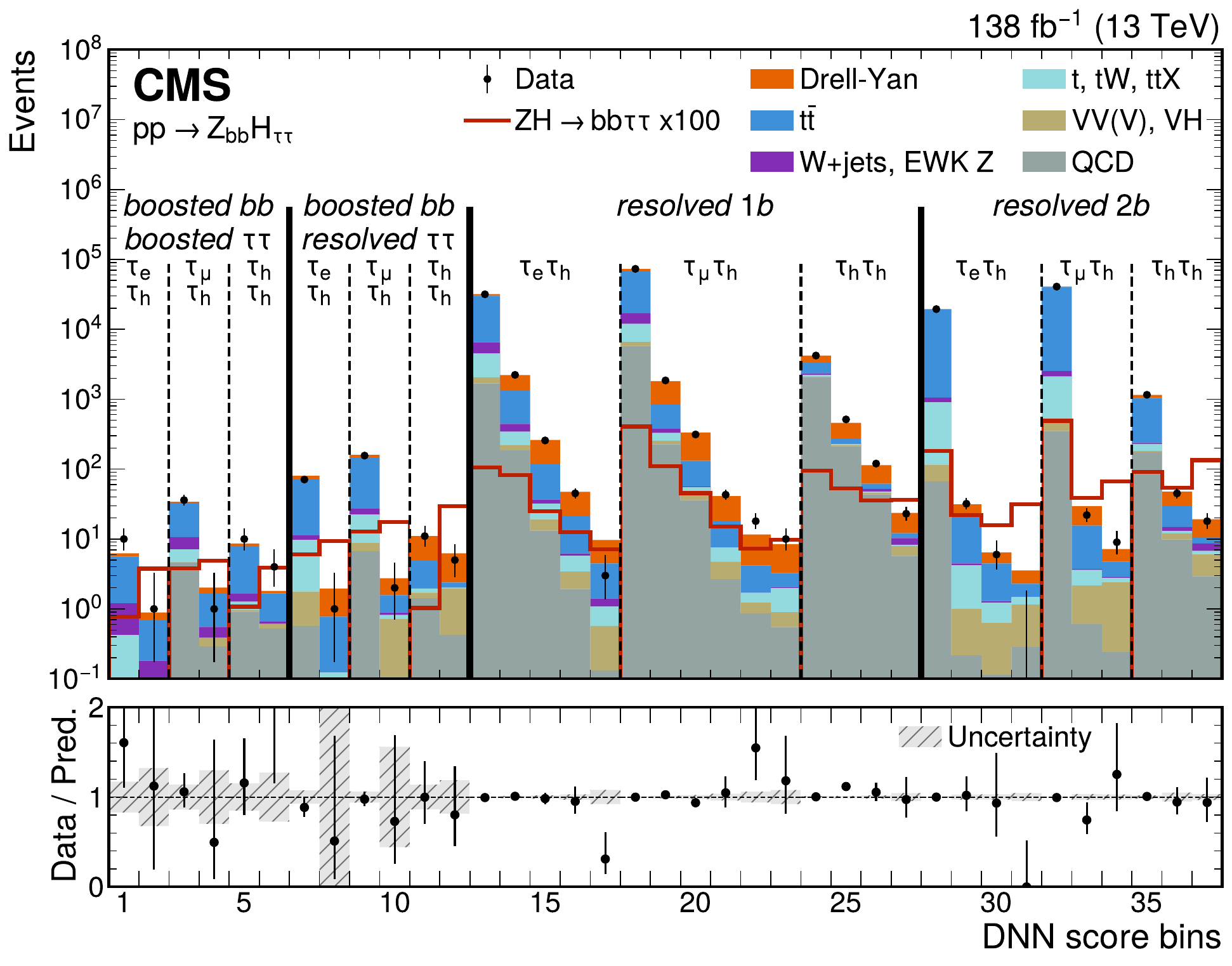}
	\includegraphics[width=0.75\textwidth]{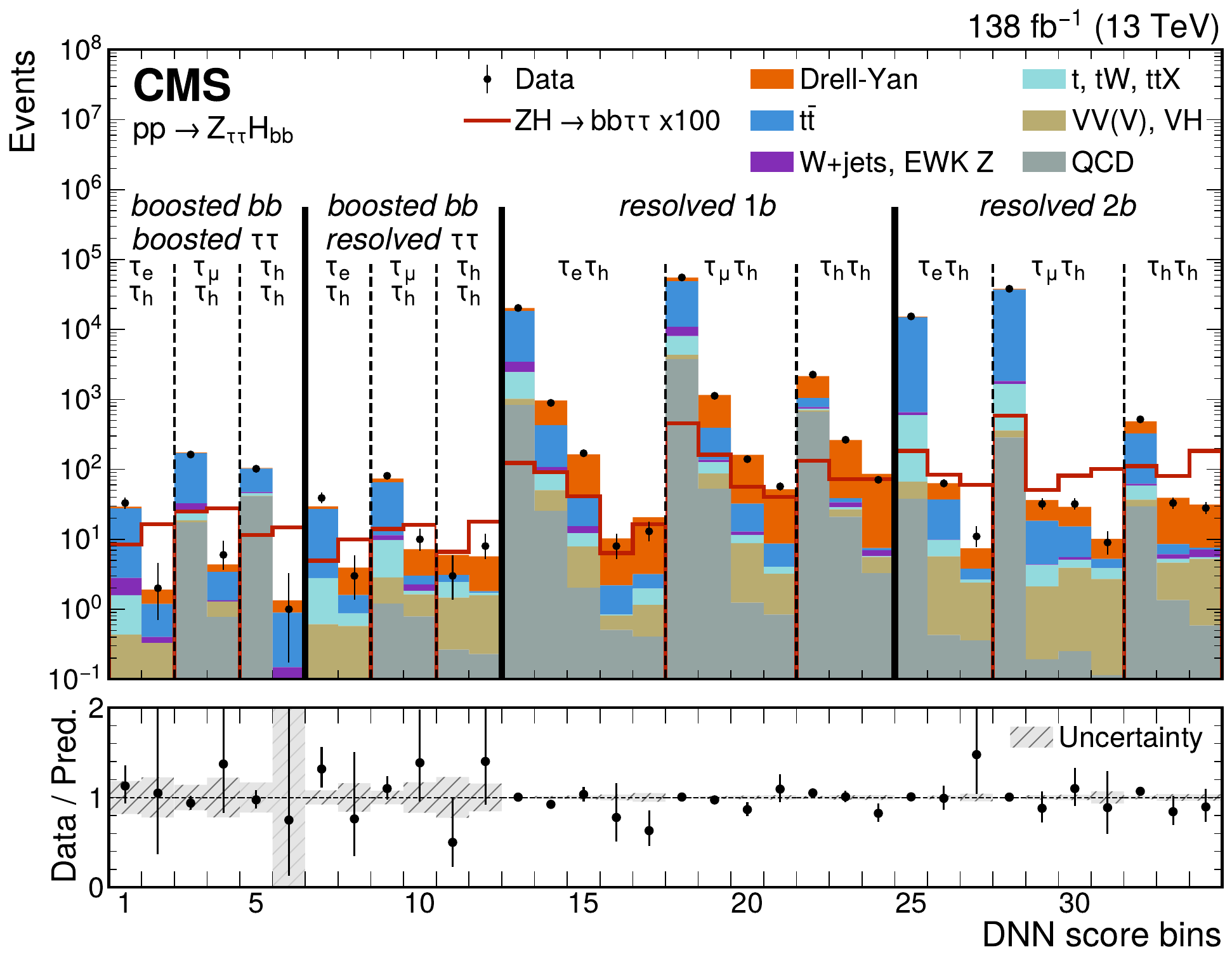}
	\caption{Distributions of the output of the DNN used for SM \ZHbbtt signal extraction. Upper: \zbbhtt signal region. Lower: \ztthbb signal region. The lower panels show the ratio of the data to the background prediction with uncertainties from a background-only fit to the observed data.}
	\label{fig:DNNdistributionsNonResZH}
\end{figure}

The signal region distributions of the nonresonant DNN used for the SM measurements are presented in Figs.~\ref{fig:DNNdistributionsNonResZZ} and~\ref{fig:DNNdistributionsNonResZH} for the \ZZ and \ZH processes, respectively. The bin width varies and is chosen to maximize the expected sensitivity while keeping uncertainties from limited simulated sample sizes at an acceptable level.


\begin{figure}[htbp!]
	\centering
	\includegraphics[width=0.71\textwidth]{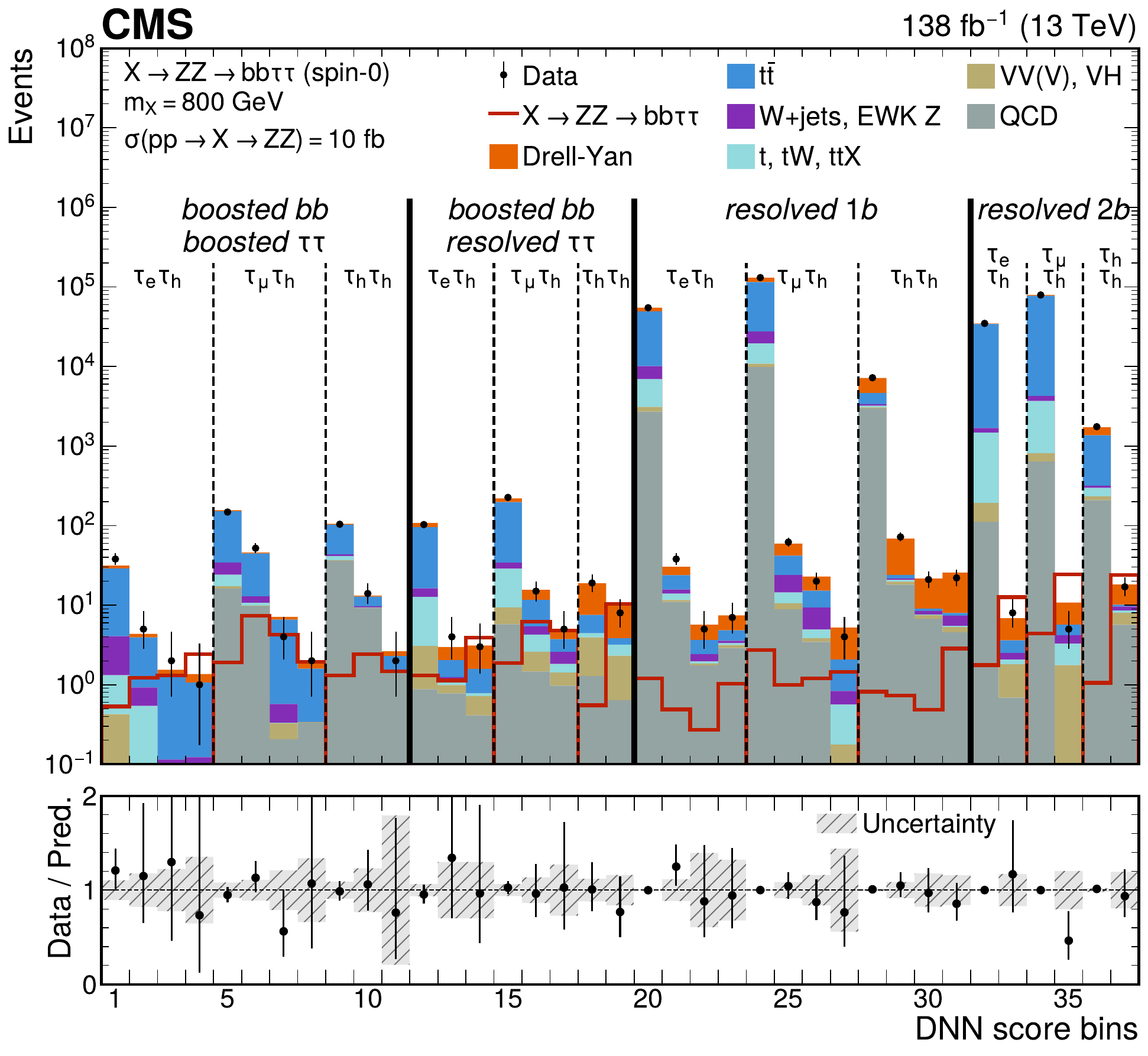}
   \includegraphics[width=0.65\textwidth]{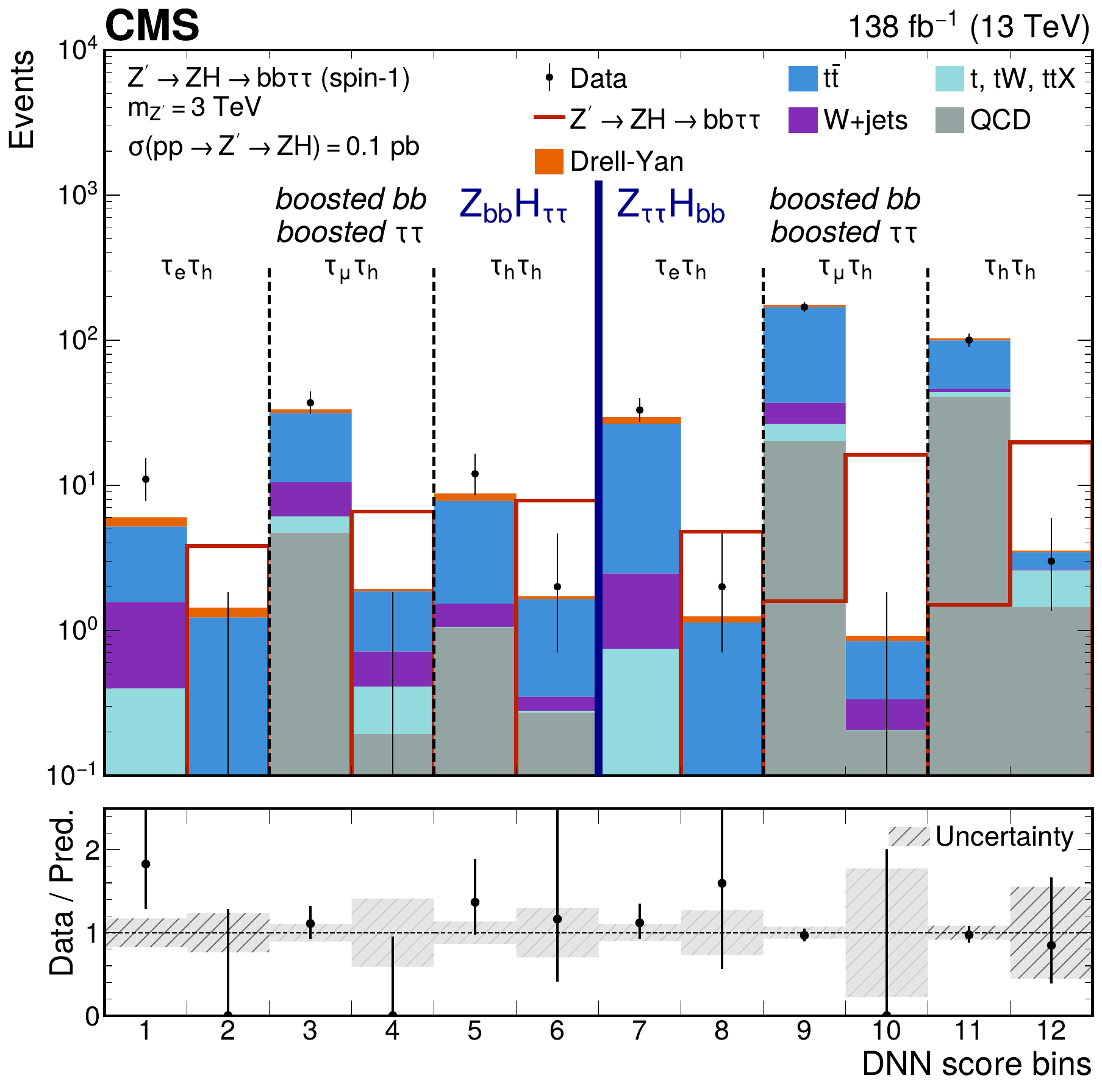}
	\caption{Example distributions of the output of the resonant DNN used for signal extraction. Upper: \XZZbbtt with $m_\PX=800\GeV$. Lower: \ZprimeZH with $m_\PZpr=3\TeV$, showing the \zbbhtt and \ztthbb signal regions in the left and right parts, respectively. The lower panels show the ratio of the data to the SM prediction with uncertainties from a background-only fit to the observed data.}
	\label{fig:DNNdistributionsRes}
\end{figure}

Representative distributions of the parameterized DNN score used for the resonant searches are shown in Fig.~\ref{fig:DNNdistributionsRes}. The parameterized DNN provides good discrimination against the background over a wide range of resonance masses. For each mass hypothesis, only regions with a signal acceptance ${>}0.01\%$ are included in the fit, as certain topologies have negligible acceptance in specific mass ranges (\eg resolved categories at very high masses).

\section{Systematic uncertainties}\label{Sec:Systematics}
Systematic uncertainties arise from model assumptions, analysis methods, and the experimental setup. Uncertainties affecting both the normalization and shape of the signal and background distributions are considered.
The integrated luminosities for the 2016, 2017, and 2018 data-taking years have individual uncertainties of 1.2--2.5\%~\cite{CMS-LUM-17-003,CMS-PAS-LUM-20-001}, while the overall uncertainty for the 2016--2018 period is 1.6\%.

For the reconstruction of resolved \tauh candidates, several corrections and uncertainties are applied. Identification uncertainties for \tauh candidates are in the 5--20\% range. Energy scale uncertainties are applied to genuine \tauh candidates as a function of \pt and decay mode and range from 0.6 to 0.8\%~\cite{TauPerfRun2}. Identification and energy scale corrections are also applied to electrons and muons misidentified as \tauh candidates; these corrections are measured using the $\PZ\to\taumu\tauh$ and $\PWst\to\tauh\PGn$ processes.
For the reconstruction of boosted \tauh candidates, a 10\% uncertainty in the reconstruction and ID efficiency is applied per \tauh candidate, measured using $\PZ\to\taumu\tauh$ events~\cite{DeepBoostedTauDPS}, as well as a 3\% uncertainty in the energy scale.

Jet energy scale and resolution uncertainties are considered as functions of the data-taking period and jet flavor. They affect both AK4 and AK8 jets and are propagated to the \ptmiss calculation.
Uncertainties in \PQb tagging efficiency, accounting for contamination from light-quark, gluon and charm jets in heavy-flavor regions, are applied for AK4 jets~\cite{BTV-16-002,BTVDeepJetDPS}. For AK8 jets, uncertainties in the efficiency of the \ParticleNet $\PQb\PQb$ tagger are considered separately for signal-like processes~\cite{BoostedBBJetIDPerf} and for the dominant backgrounds.

Additional reconstruction uncertainties include: trigger efficiency uncertainties; uncertainties in electron and muon isolation and ID efficiencies, as well as energy scale; uncertainties in the ECAL and muon systems prefiring corrections (ECAL prefiring corrections are applied to 2016--2017 simulated samples, whereas muon prefiring corrections are applied for all years); uncertainties in the efficiency of the pileup jet ID algorithm~\cite{Sirunyan:2020foa}; and the pileup reweighting uncertainty, assessed by varying the total inelastic cross section used to derive pileup distributions in simulation by ${\pm}4.6\%$~\cite{CMS:2018mlc}.

Uncertainties due to the limited size of simulated Monte Carlo samples are included in the likelihood, following Refs.~\cite{BarlowBeeston,ConwayLikelihood}.

Uncertainties in PDFs and strong coupling \alpS for signal and background processes are estimated by varying the PDF set or the value of \alpS by one standard deviation with respect to the nominal value~\cite{PDFLHC}. Renormalization and factorization scale uncertainties are estimated by simultaneously varying these two scales between 0.5 and 2 times their nominal values, while keeping the ratio of renormalization to factorization scale between 0.5 and 2. The impact of these uncertainties is small for most processes, except for the DY background where the uncertainty ranges from 12 to 60\% depending on the category, owing to the large renormalization scale uncertainty in \PZ boson production in association with one or two \PQb quarks.

An uncertainty in the yield and shape of the QCD background is assigned, derived from the limited number of data and simulated events available in the control regions used by the ABCD method.

For the SM measurements and the resonant analyses at low masses ($m_{\PX/\PZpr}<1\TeV$), the dominant systematic uncertainties are theoretical uncertainties in the DY background prediction, uncertainties due to the limited size of simulated samples, and uncertainties related to AK4 \PQb tagging efficiencies. In the resonant analyses at high masses ($m_{\PX/\PZpr}>1\TeV$), the dominant systematic uncertainties are related to the efficiency of the \ParticleNet $\PX\to\PQb\PQb$ tagger, the limited size of simulated samples, and theoretical uncertainties in the \ttbar background prediction.

\section{Results}\label{Sec:Results}
A binned maximum likelihood fit of the predicted DNN distribution to that in the data is simultaneously performed in the four categories described in Section~\ref{Subsec:Categorization} (\resb, \resbb, \catBoostedHPS, and \catBoostedTaus), the three channels (\eleth, \muth, and \thth) and all data-taking periods.
The systematic uncertainties described in Section~\ref{Sec:Systematics} are introduced as nuisance parameters in the likelihood function~\cite{HiggsCombinationProcedure2011}.

\begin{figure}[htbp!]
	\centering
	\includegraphics[width=0.49\textwidth]{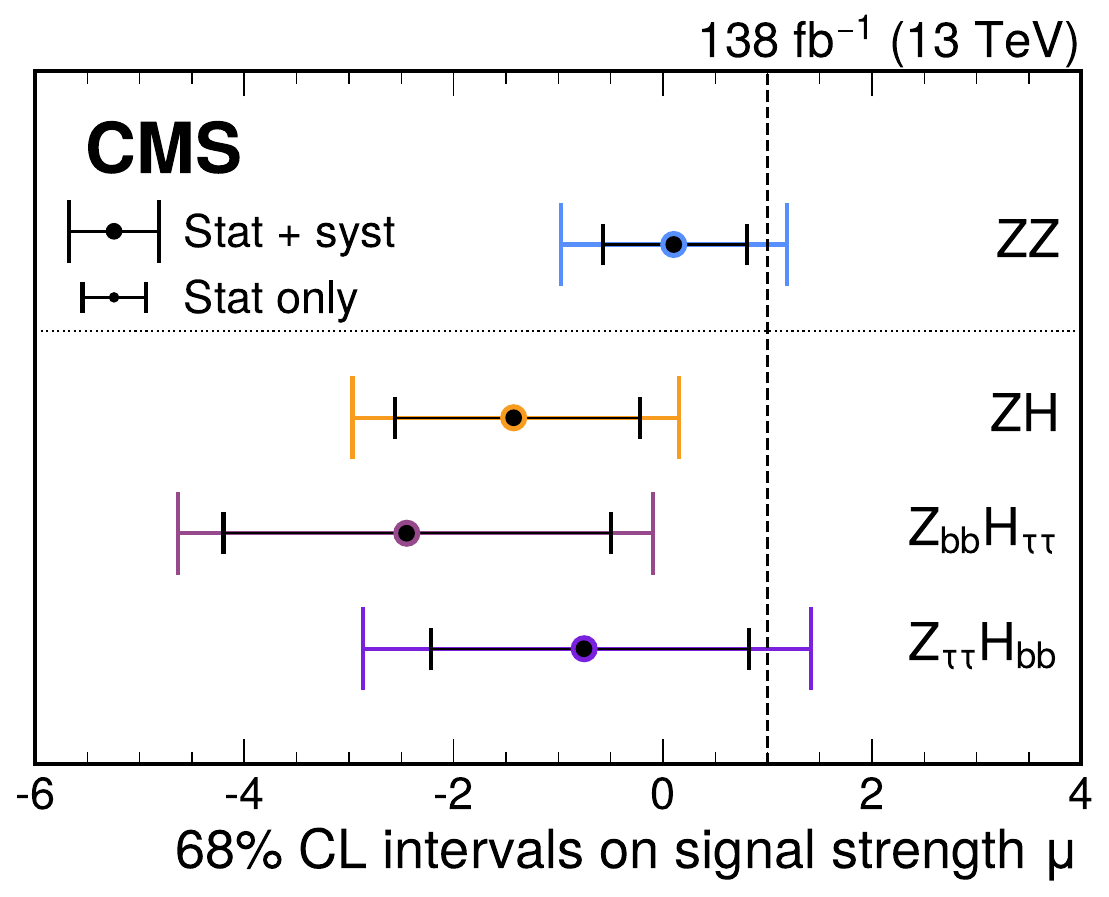}
    \includegraphics[width=0.49\textwidth]{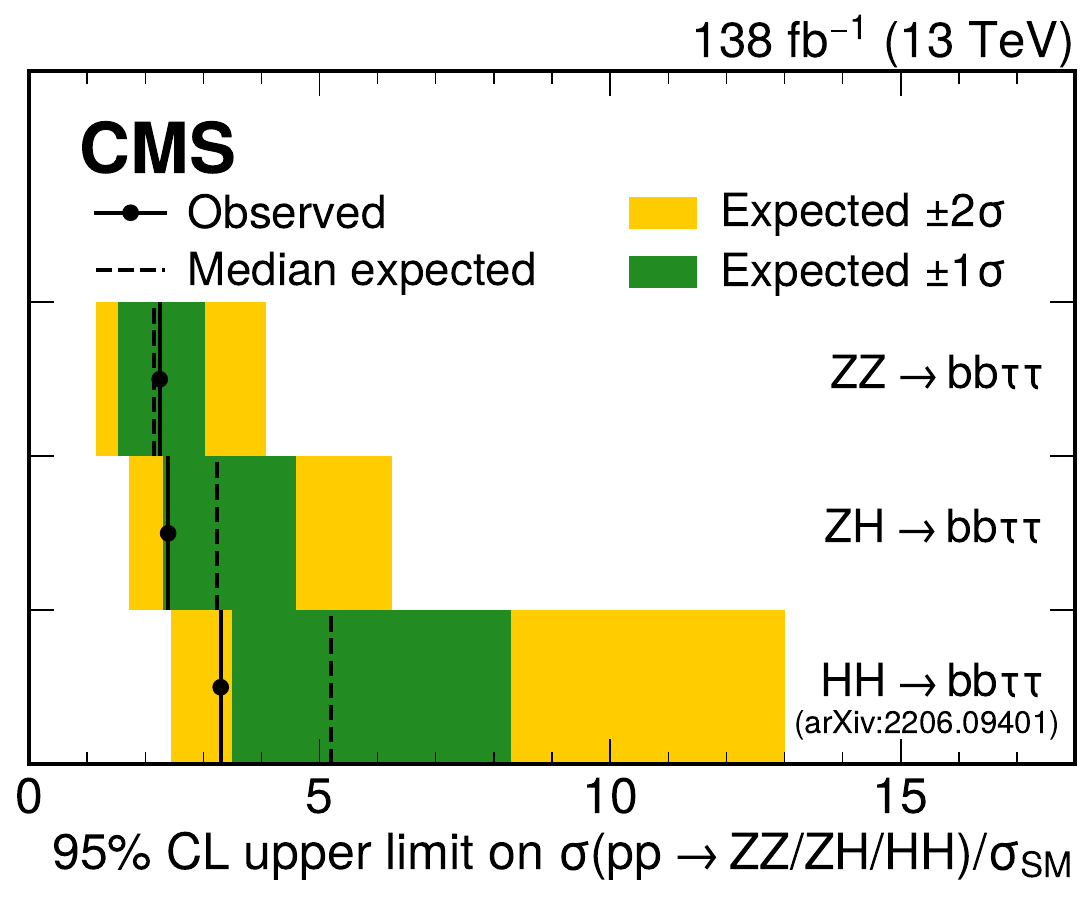}
	\caption{Results for the SM \ZZbbtt and \ZHbbtt processes. Left: best fit signal strengths and approximate 68\% \CL intervals, obtained from a profile likelihood fit, as described in Ref.~\cite{LikelihoodScanInterval}. Right: upper limits at 95\% \CL on the ratio of production cross section and the SM expectation for the \ZZ and \ZH processes, as well as those from the \HHbbtt analysis on the same data set for comparison~\cite{CMSRun2HHbbtt}.}
	\label{fig:LikelihoodScanSM}
\end{figure}

Results for the SM processes \ZZbbtt and \ZHbbtt are presented in Fig.~\ref{fig:LikelihoodScanSM}. The best fit value of the signal strength $\mu$, where $\mu$ denotes the ratio of the measured cross section to the SM theoretical cross section, and an approximate confidence interval at 68\% confidence level (\CL) are extracted following the procedure described in Section~3.2 of Ref.~\cite{LikelihoodScanInterval}.
The results are obtained using the CMS statistical analysis tool \textsc{Combine}~\cite{CMS:2024onh}, based on the \textsc{RooFit}~\cite{Verkerke:2003ir} and \textsc{RooStats}~\cite{Moneta:2010pm} frameworks.
The best fit signal strength is $\mu_{\ZZ} = 0.10^{+0.70}_{-0.68} \stat {}^{+0.83}_{-0.84} \syst$ for \ZZ, and $\mu_{\ZH} = -1.4^{+1.2}_{-1.1} \stat \pm 1.0 \syst$ for \ZH. The results are compatible with SM predictions.

The most sensitive category for the SM measurements is \resbb, followed by the boosted categories. The most sensitive decay channel is \thth, owing to its larger branching fraction and lower background, followed by \muth, because of its higher efficiency.

Additionally, upper limits at 95\% \CL are set on the product of production cross section and branching fraction of the SM \ZZ and \ZH processes. The \CLs criterion~\cite{CLS1,CLS2} is used in the limit-setting procedure. The profile likelihood ratio, modified for upper limits, is used as the test statistic~\cite{HiggsCombinationProcedure2011}, and the asymptotic approximation~\cite{Cowan:2010js} is used to estimate the test statistic distribution.
The observed (expected) upper limits are 2.3 (2.1) times the SM expectation for \ZZ, and 2.4 (3.2) times for \ZH, respectively. They are shown in Fig.~\ref{fig:LikelihoodScanSM} (right).
The \HHbbtt analysis obtained observed (expected) upper limits of 3.3 (5.2) times the SM expectation on the same data set~\cite{CMSRun2HHbbtt}.
Compared to \HHbbtt, the \ZZZHbbtt processes have larger cross sections, resulting in more stringent expected upper limits. However, increased background contributions in the \ZZ signal region and reduced selection efficiencies due to lower-\pt final-state objects lead to sensitivities that remain below those suggested by the cross section ratios alone.
The comparison with \HHbbtt indicates that the methods used in this analysis provide a sensitive probe of the \bbtt final state and can help establish analysis techniques relevant for future \HH measurements.

As the ongoing 2022--2026 (Run~3) data-taking and the future HL-LHC program will provide substantially larger data sets, the current results can be extrapolated to higher integrated luminosities.
For these projections, the baseline assumption is that the detector performance and analysis strategy remain unchanged with respect to the current results. The first considered scenario (denoted S2) assumes reduced systematic uncertainties and negligible MC statistical uncertainties, following Ref.~\cite{HLLHCprojections}. In addition, an optimistic scenario is considered in which systematic uncertainties are neglected, providing an upper-bound of the sensitivity.

\begin{figure}[htbp]
    \centering
    \includegraphics[width=0.55\textwidth]{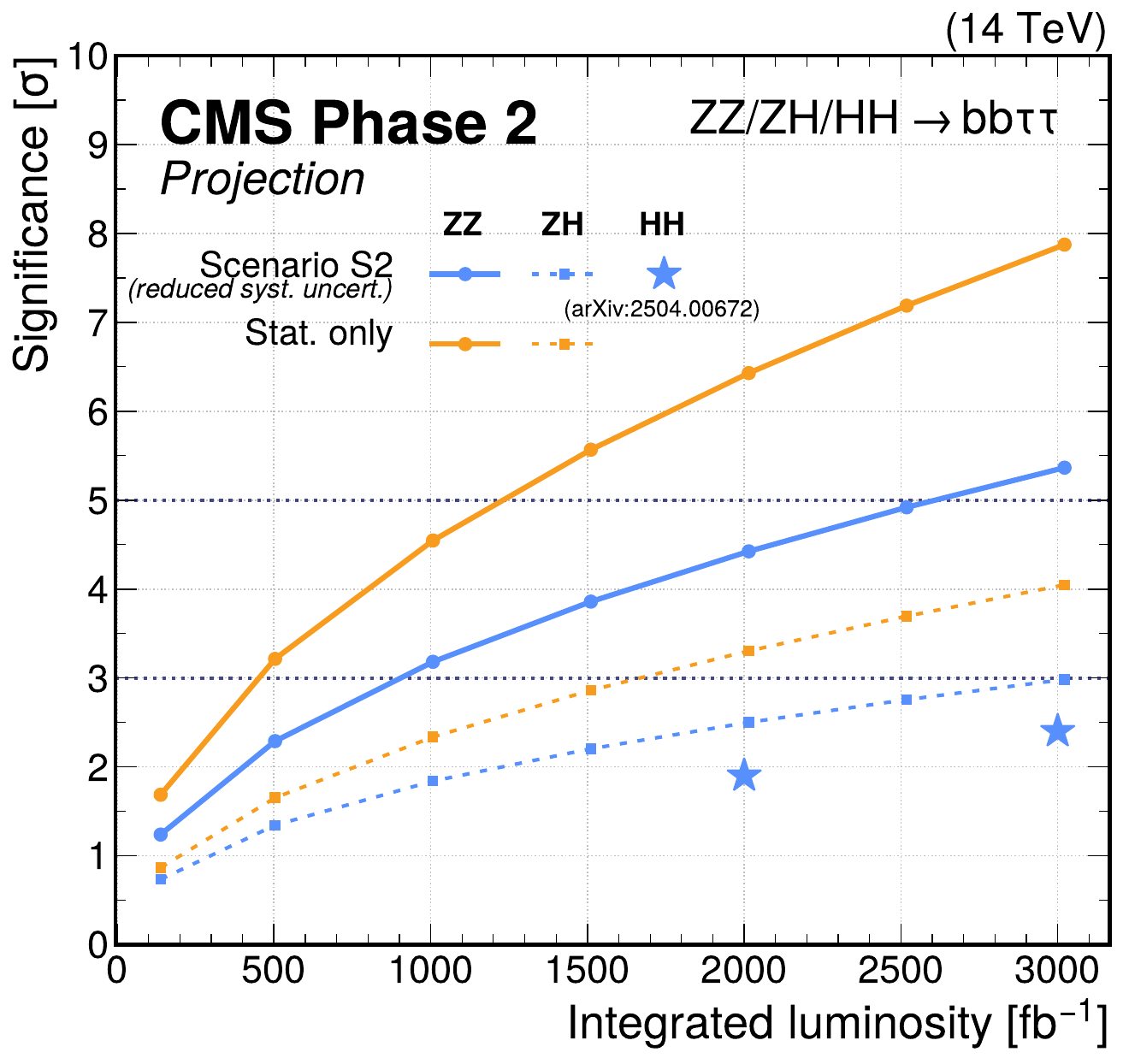}
    \caption{Projected significance~\cite{pvalue} for SM \ZZbbtt and \ZHbbtt processes as a function of integrated luminosity at the HL-LHC with the CMS Phase~2 detector upgrade~\cite{PrimaryVertexFromPhase2TDR}. Two scenarios for systematic uncertainties are shown: the S2 scenario assumes reduced systematic uncertainties, as described in Ref.~\cite{HLLHCprojections}, and the second scenario neglects systematic uncertainties. Projections for \HHbbtt from Ref.~\cite{HLLHCprojections} are also shown for comparison. The thresholds required for evidence ($3\sigma$) and observation ($5\sigma$) are indicated.
    }
    \label{fig:Prospects}
\end{figure}

The projected significances~\cite{pvalue} for the \ZZZHbbtt analyses as a function of the integrated luminosity are illustrated in Fig.~\ref{fig:Prospects}.

The results suggest that evidence for the \ZZbbtt process may be reached with approximately 1\abinv of integrated luminosity. Observation of the process would require more than 2.5\abinv, corresponding to slightly less than the expected integrated luminosity per experiment at the end of the HL-LHC program. The projections improve significantly when systematic uncertainties are neglected, highlighting the need for further refinements in object reconstruction, detector performance, and background modeling.
Compared to the \HHbbtt analysis, evidence for \ZZbbtt is expected with substantially less integrated luminosity, motivating the use of the \ZZbbtt process as a benchmark to assess analysis techniques for the \bbtt final state.
The outlook for the \ZHbbtt process is comparatively more challenging. Even with the anticipated 3\abinv HL-LHC data set, an observation will likely remain out of reach without substantial reductions in systematic uncertainties and further improvements in analysis methods.

\begin{figure*}[th]
	\centering
	\includegraphics[width=0.49\textwidth]{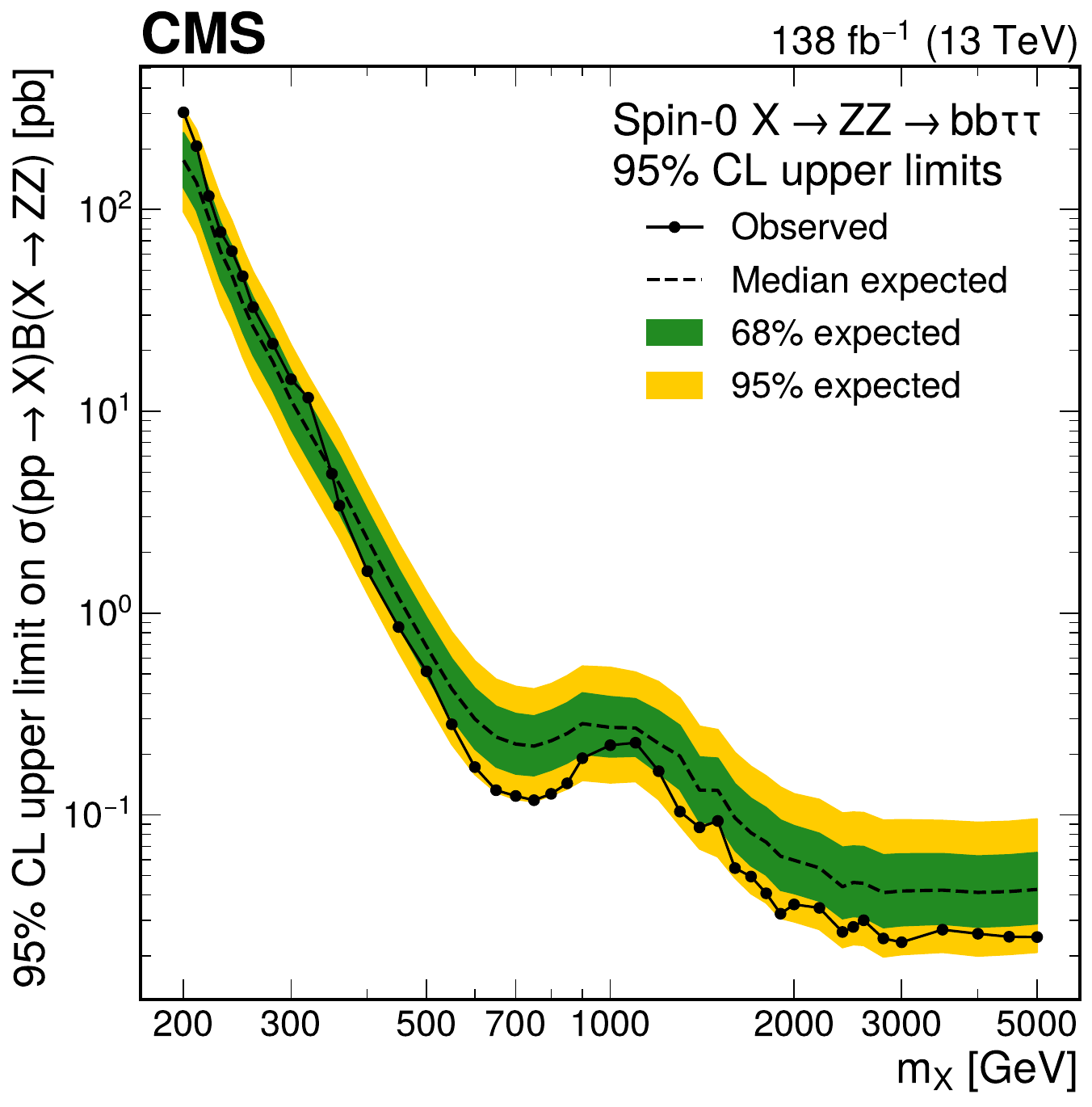}
    \includegraphics[width=0.49\textwidth]{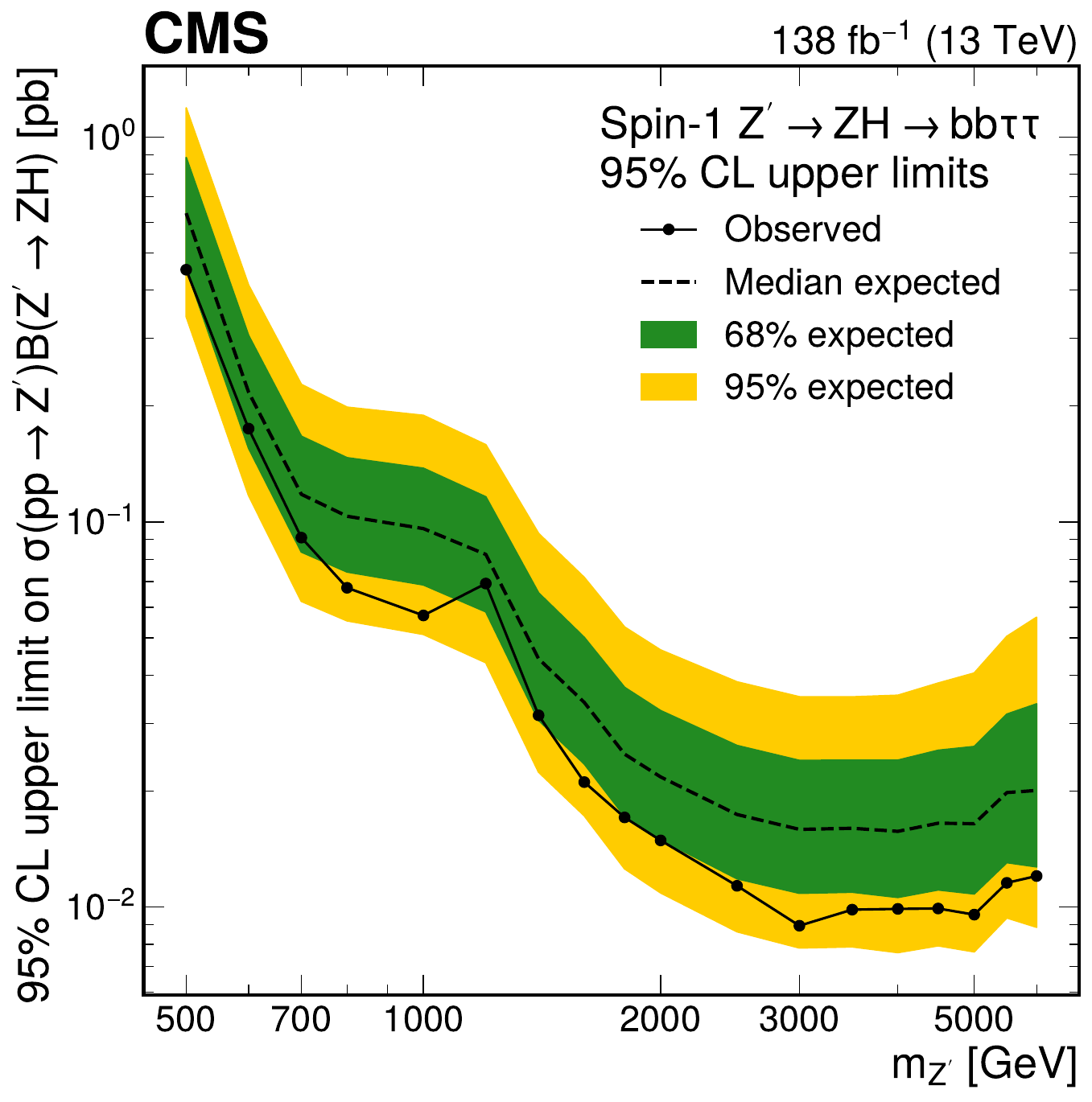}
	\caption{Upper limits on the production cross section for the \XZZ process (left, spin-0 resonance) and the \ZprimeZH process (right, spin-1 resonance), obtained under the narrow-width approximation.}
	\label{fig:UpperLimits}
\end{figure*}

Upper limits at 95\% \CL are also set on the production cross section for resonant \ZZ and \ZH production under the narrow-width approximation (via spin-0 and spin-1 resonances, respectively), and are shown in Fig.~\ref{fig:UpperLimits}.

At low resonance masses, the sensitivity is driven by the resolved categories, with weak exclusion power due to large backgrounds and low trigger and reconstruction efficiencies. The \muth decay channel is the most sensitive in this region. As the resonance mass increases, the \thth channel becomes the most sensitive. The transition between resolved and boosted categories takes place around 1\TeV, leading to some loss of sensitivity in this range. At very high masses, the \muth channel becomes the most sensitive due to its higher reconstruction efficiency.

The observed limits at high masses are slightly stronger than expected because of downward fluctuations of the data relative to the background prediction, within the uncertainties, in several of the most sensitive bins of the final discriminant.
Since the \ZZ and \ZH analyses share overlapping regions of phase space, the fluctuations are expected to be present in both spin-0 and spin-1 resonant search results.
\newpage

\section{Summary}\label{Sec:Summary}
A study of \ZZ and \ZH production in the \bbtt final state has been presented based on proton-proton collisions collected by the CMS experiment at the CERN LHC at $\sqrt{s}=13\TeV$, corresponding to an integrated luminosity of 138\fbinv.

Both nonresonant and resonant production mechanisms are investigated. In the nonresonant interpretation, the production of \ZZbbtt and \ZHbbtt is studied, and the results are found consistent with standard model predictions.
Projections of the sensitivity of the \ZZ and \ZH measurements that could be achieved at the high-luminosity LHC are presented. These projections indicate that the \ZZbbtt production is expected to reach the observation threshold earlier than that of the \HHbbtt search, demonstrating for the first time the use of the \ZZbbtt process as a standard candle to assess the analysis methods for the \HH search in the \bbtt final state.

In the resonant interpretation, searches are performed for heavy spin-0 resonances that decay into \ZZ and for spin-1 resonances that decay into \ZH, with resonance masses between 200\GeV and 6\TeV.
No evidence for a signal is observed, and upper limits are set on the production cross sections.
These are the first measurements to probe the \ZZZHbbtt processes.

\bibliography{auto_generated}
\cleardoublepage \appendix\section{The CMS Collaboration \label{app:collab}}\begin{sloppypar}\hyphenpenalty=5000\widowpenalty=500\clubpenalty=5000\cmsinstitute{Yerevan Physics Institute, Yerevan, Armenia}
{\tolerance=6000
A.~Gevorgyan\cmsorcid{0000-0003-2751-9489}, A.~Hayrapetyan, V.~Makarenko\cmsorcid{0000-0002-8406-8605}, A.~Tumasyan\cmsAuthorMark{1}\cmsorcid{0009-0000-0684-6742}
\par}
\cmsinstitute{Marietta Blau Institute for Particle Physics, Vienna, Austria}
{\tolerance=6000
W.~Adam\cmsorcid{0000-0001-9099-4341}, L.~Benato\cmsorcid{0000-0001-5135-7489}, T.~Bergauer\cmsorcid{0000-0002-5786-0293}, M.~Dragicevic\cmsorcid{0000-0003-1967-6783}, S.~Gundacker\cmsorcid{0000-0003-2087-3266}, A.K.~Guven\cmsorcid{0009-0004-5670-5138}, P.S.~Hussain\cmsorcid{0000-0002-4825-5278}, M.~Jeitler\cmsAuthorMark{2}\cmsorcid{0000-0002-5141-9560}, N.~Krammer\cmsorcid{0000-0002-0548-0985}, A.~Li\cmsorcid{0000-0002-4547-116X}, D.~Liko\cmsorcid{0000-0002-3380-473X}, M.~Matthewman, J.~Schieck\cmsAuthorMark{2}\cmsorcid{0000-0002-1058-8093}, R.~Sch\"{o}fbeck\cmsAuthorMark{2}\cmsorcid{0000-0002-2332-8784}, M.~Shooshtari\cmsorcid{0009-0004-8882-4887}, M.~Sonawane\cmsorcid{0000-0003-0510-7010}, N.~Van~Den~Bossche\cmsorcid{0000-0003-2973-4991}, W.~Waltenberger\cmsorcid{0000-0002-6215-7228}, C.E.~Wulz\cmsAuthorMark{2}\cmsorcid{0000-0001-9226-5812}
% ...
\par}
$^{1}$Also at Yerevan State University, Yerevan, Armenia\\
$^{2}$Also at Technische Universit\"{a}t Wien, Vienna, Austria\\
% ...
\end{sloppypar}
\end{document}